\documentclass[final,3p,times,onecolumn, sort&compress]{elsarticle}

\usepackage{lineno,hyperref}
\modulolinenumbers[5]

\usepackage{amsmath}
\usepackage{amsfonts,color}
\usepackage{amssymb,float}
\usepackage{mathtools}
\usepackage[utf8]{inputenc}
\usepackage{graphicx}
\usepackage{wasysym}
\usepackage{tabularx}
\usepackage{accents}
\usepackage{graphicx}
\usepackage{color, colortbl}
\usepackage[dvipsnames]{xcolor}
\usepackage{hyperref}
\usepackage{ulem}

\hypersetup{
    colorlinks=true,
    linkcolor=blue,
    filecolor=magenta,      
    citecolor=red
}

\usepackage{widetext}

\usepackage{subfigure} 
\usepackage{hyperref} 
\usepackage{multirow}

\definecolor{Gray}{gray}{0.9}
\definecolor{LightCyan}{rgb}{0.88,1,1}

\newcommand{\be}{\begin{equation}}
\newcommand{\ee}{\end{equation}}
\newcommand{\bea}{\begin{eqnarray}}
\newcommand{\eea}{\end{eqnarray}}

\journal{}

\begin{document}

\begin{frontmatter}

\title{Dark energy by natural evolution: \\ Constraining dark energy using Approximate Bayesian Computation}


\author[a]{Reginald Christian Bernardo\corref{mycorrespondingauthor}}
\cortext[mycorrespondingauthor]{Corresponding author}
\ead{rbernardo@gate.sinica.edu.tw}

\author[b]{Daniela Grand\'on}
\ead{daniela.grandon@ug.uchile.cl}

\author[c,d]{Jackson Levi Said}
\ead{jackson.said@um.edu.mt}

\author[e]{V\'ictor H. C\'ardenas}
\ead{victor.cardenas@uv.cl}

\address[a]{Institute of Physics, Academia Sinica, Taipei 11529, Taiwan}
\address[b]{Grupo de Cosmolog\'ia y Astrof\'isica Te\'orica, Departamento de F\'isica, FCFM, Universidad de Chile, Blanco Encalada 2008, Santiago, Chile}
\address[c]{Institute of Space Sciences and Astronomy, University of Malta, MSD 2080, Malta}
\address[d]{Department of Physics, University of Malta, MSD 2080, Malta}
\address[e]{Instituto de F\'isica y Astronom\'ia, Universidad de Valpara\'iso, Av. Gran Bretaña 1111, Valparaiso, Chile}

\begin{abstract}
We look at dark energy from a biology inspired viewpoint by means of the Approximate Bayesian Computation (ABC) and late time cosmological observations. We find that dynamical dark energy comes out on top, or in the ABC language naturally selected, over the standard $\Lambda$CDM cosmological scenario. We confirm this conclusion is robust to whether baryon acoustic oscillations and Hubble constant priors are considered. Our results show that the algorithm prefers low values of the Hubble constant, consistent or at least a few standard deviation away from the cosmic microwave background estimate, regardless of the priors taken initially in each model. This supports the result of the traditional MCMC analysis and could be viewed as strengthening evidence for dynamical dark energy being a more favorable model of late time cosmology. 
\end{abstract}


\end{frontmatter}




\section{Introduction} \label{sec:intro}

The standard model of cosmology, or $\Lambda$CDM, is the most widely accepted explanation of observations both in cosmology and astrophysics \cite{Peebles:2002gy,Copeland:2006wr}. It describes the late--time accelerated expansion \cite{Riess:1998cb,Perlmutter:1998np} through a cosmological constant $\Lambda$ which appears with general relativity (GR) to model gravitational interactions. Additionally, the appearance of cold dark matter (CDM) acts to stabilize the dynamics of galaxies \cite{Baudis:2016qwx,Bertone:2004pz} together with their clusters. With the inclusion of inflation \cite{Guth:1980zm,Linde:1981mu}, $\Lambda$CDM can produce a continuous evolution from the big bang to current times. However, fundamental problem have persisted in the use of the cosmological constant in $\Lambda$CDM \cite{RevModPhys.61.1}, and the direct detection of CDM particles remains elusive \cite{LUX:2016ggv,Gaitskell:2004gd}. More recently, a growing tension has also emerged from the observational sector in form of tensions in the measurement of the Hubble constant using local source measurements \cite{Riess:2019cxk,Wong:2019kwg,Freedman:2021ahq} and those using early Universe data in conjunction with $\Lambda$CDM to predict the value of the Hubble constant \cite{DES:2017txv,Aghanim:2018eyx}. This cosmological tension appears to be consistent across most analyses \cite{DiValentino:2020vhf,DiValentino:2020zio,DiValentino:2020vvd,Staicova:2021ajb,Vagnozzi:2019ezj,Smith:2022iax,Simon:2022adh,Smith:2022hwi,Poulin:2021bjr,Schoneberg:2021qvd,Smith:2020rxx,Murgia:2020ryi,Smith:2019ihp,Poulin:2018cxd,Poulin:2018zxs,DiValentino:2021rjj,Yang:2021eud,DiValentino:2020naf,Yang:2020zuk,DiValentino:2019jae,DiValentino:2019ffd,Pan:2019gop,DiValentino:2019exe,Yang:2018qmz,Yang:2018euj,Cyr-Racine:2021oal,Vagnozzi:2021gjh,Brinckmann:2020bcn,RoyChoudhury:2020dmd,Jedamzik:2020krr,Ye:2020btb,Vagnozzi:2019ezj,Lin:2019qug,Agrawal:2019lmo,Kreisch:2019yzn}.

The fundamental physics and observational challenges to $\Lambda$CDM has led to a reconsideration of cosmological models beyond the standard model. These range from new proposals for CDM physics with novel interactions \cite{Feng:2010gw,Dodelson:1993je,Joyce:2014kja,Abazajian:2012ys}, to dynamical dark energy (DE) models \cite{Copeland:2006wr,Benisty:2021gde,Benisty:2020otr,Bamba:2012cp}, as well as modifications to GR \cite{Clifton:2011jh,CANTATA:2021ktz,Bahamonde:2021gfp,AlvesBatista:2021gzc,Addazi:2021xuf,Capozziello:2011et}. By and large, these models incorporate additional layers of complexity into either the matter or gravitational sectors of the field equations, which may have drastic impacts on the computational time for cosmological analyses. This fact is made all the worse by the rapid increases in the amount of observational data available \cite{Planck:2018vyg,ACT:2020gnv,eBOSS:2020yzd,DES:2017txv,Riess:2021jrx,Scolnic:2021amr}. Within this context, there have been suggestions in the literature that a better approach may be to consider phenomenological motivated models \cite{Cardenas:2014jya}. Taking this approach, observational data seems to point away from $\Lambda$CDM as shown in Ref.~\cite{Grandon:2021nls} where a parametrized dynamical dark energy approach was used to make this assessment. Later on, in Ref.~\cite{Bernardo:2021cxi} a further analysis was performed with lower uncertainties and considering the Gaussian Processes (GP) pointing to potentially new physics on cosmic scales.

GP \cite{10.5555/1162254} is a useful supervised learning approach which can reconstruct cosmological parameters together with their uncertainties consistently across most data sets. Here, a covariance function accounts for the correlations between the data points within a particular data set. The covariance, or kernel function, does this through a number of statistical hyperparameters, which are not related to the physics at play within the data itself. These hyperparameters characterize the amplitude and length scales on which the correlations can occur. At the level of the Hubble diagram, this has been used in various scenarios \cite{OColgain:2021pyh,Busti:2014aoa,Busti:2014dua,Seikel:2013fda,Yahya:2013xma,2012JCAP...06..036S,Shafieloo:2012ht,Bernardo:2021qhu,Ruiz-Zapatero:2022zpx,Benisty:2022psx,Briffa:2020qli,Ren:2022aeo,Cai:2019bdh}, together with comparisons with other approaches to cosmology independent reconstructions \cite{Escamilla-Rivera:2021rbe}. This approach has also been applied to observations characterizing the large scale structure of the Universe \cite{Benisty:2020kdt,LeviSaid:2021yat,Reyes:2022exv}. However, GPs suffer from the problems of overfitting at low redshift as well as selection problems in the choice of the kernel function. In Ref.~\cite{Bernardo:2021mfs}, the problem of selecting a kernel is probed by using an intriguing application of genetic algorithms, which in conjunction with GP can be used to reconstruct data without an over-heavy reliance on the choice of kernel. Here, the kernel selection problem is confronted using the approximate Bayesian computation (ABC) approach which is a likelihood-free inferencing algorithm that is inspired from natural evolution. For each iteration or generation of a family of kernel functions, a comparison with data is made, analogous to the likelihood calculation in regular Markov Chain Monte Carlo (MCMC) analyses. Given a data set $D$ and model parameter $\theta$ with posterior distribution $P(\theta \vert D)$, the purpose of ABC is to approximate this distribution, i.e. $P(\theta\vert D) \propto {\cal L}(D\vert \theta)P(\theta)$, where ${\cal L}(D \vert \theta)$ is the likelihood of $\theta$ for data $D$ \cite{10.3389/fbuil.2017.00052}. By interfacing ABC with sequential Monte Carlo sampling (SMC), an ABC-SMC algorithm can be competitive with the regular MCMC approach \cite{Toni_2008, 2009arXiv0911.1705T, 2009arXiv0910.4472T}. In fact, this approach has been used in a number of scenarios in the literature \cite{Akeret:2015uha, 2013ApJ...764..116W, 2017A&C....19...16J, Ishida:2015wla}.

In the context of the GP kernel selection problem, ABC-SMC offers a useful approach to remove the problem of arbitrarily selecting a kernel function, and automate the search for the best kernel that approximates the observational data. The model space from which kernels can be selected or combined together is potentially infinite, making the problem potentially intractable analytically. ABC can tackle this problem by determining the best combination of kernels that approximates the observational data using ABC-SMC \cite{10.5555/534133}. ABC is partially modelled by the principles of biological evolution where a form of natural selection together with generational mutations allows for an efficient process by which consecutive generations can be produced that approximate data progressively well. We use ABC to eliminate the kernel selection issue by finding the `fittest' kernels. In cosmology, such genetic algorithms have been shown to be competitive with other methods \cite{2012arXiv1202.1643R,Bogdanos:2009ib,Arjona:2019fwb,2012JCAP...11..033N}. Saying that, other approaches within the toolkit of machine learning have also been applied to observational data such as neural network systems \cite{Escamilla-Rivera:2021vyw,Aljaf:2022fbk,Lemos:2022kua,Dialektopoulos:2021wde,Mukherjee:2022yyq,aggarwal2018neural,Wang:2020sxl,Gomez-Vargas:2021zyl,Grandon:2022gdr,Gunther:2022pto,Manrique-Yus:2019hqc}, among others.

In this work, we wish to show how DDE emerges by comparing cosmological models with observational data. We do this using this implementation of ABC. ABC-SMC is especially powerful and can be competitive with the traditional parameter estimation in cosmology, in the sense that tighter constraints are obtained and that the entire process is viewed from a model space where the different models themselves are each provided their own statistical identities. The algorithm proceeds to show which model best represents the data all while estimating the parameters buried deep within each competing model. Through this, we show how DDE naturally comes out by subjecting $\Lambda$CDM and its DDE parametric extensions (Section \ref{sec:dark_energy}) to a natural evolution test hosted by ABC. To perform this analysis we use the `pyabc' Python package \cite{10.1093/bioinformatics/bty361} which implements all the steps together in an efficient and user friendly manner. We also encourage the readers to take a look at our python notebooks, baring all the details of this work. The paper is organized as following, Sec.~\ref{sec:dark_energy} contains our introduction to the DE models, while in Sec.~\ref{sec:cosmodat}, the data sets under consideration are described. Sec.~\ref{sec:abcvsmcmc} then provides some technical details of the ABC algorithm. The results are shown in Sec.~\ref{sec:results}, followed by a summary of the main conclusions of the work in Sec.~\ref{sec:discussion}.

In addition, we also consider priors on the value of the Hubble parameter as reported in the latest SH0ES local estimate \cite{Riess:2021jrx} of $H_0^{\rm R22} = 73.30 \pm 1.04 \,{\rm km\, s}^{-1} {\rm Mpc}^{-1}$ and the Hubble constant that appears in the latest release by the Planck Collaboration \cite{Aghanim:2018eyx} with $H_0^{\rm P18} = 67.4 \pm 0.5 \,{\rm km\, s}^{-1} {\rm Mpc}^{-1}$. {In this work we have considered the reported value of the Hubble constant with the higher value for our $H_0^{\rm R22}$ since we want to test the method being proposed here as robustly as possible. This is the reported value of the Hubble constant when higher redshift data is taken into consideration.} We highly encourage the readers to play our python notebooks publicly available via \href{https://github.com/reggiebernardo/notebooks/tree/main/supp_ntbks_arxiv.xxxx.xxxxx}{GitHub} \cite{reggie_bernardo_4810864} in order to grasp the intricacies of the algorithm.

\section{Dark energy}
\label{sec:dark_energy}

We refer to `dark energy' (DE) as that optically invisible fluid that sources the observed late time cosmic acceleration. We review its canonical description, and some of its parametric extensions, including the ones we intend to use for later ABC data analysis.

We start with the $\Lambda$CDM model where DE is taken as a cosmological constant $\Lambda$ to provide a negative pressure to support cosmic acceleration. This parametrically simplest description of DE is widely considered as the canonical model as together with the CDM it provides a compelling picture of the evolution of the Universe with the fewest number of parameters that fit the observable data. However, this is challenged in numerous ways. The biggest elephant in the room is the cosmological constant problem, or rather the fine tuning of vacuum energy of at least about fifty orders of magnitude to match the observed scale of DE. There is also little motivation other than simplicity why DE should turn out to be constant in time, as it is in the canonical model, and that phenomenology always seem to to support an evolving DE. The Hubble tension, among others in cosmology today, further encourages a reexamination of the cornerstones of cosmology on which $\Lambda$CDM is based on. With these, it is important to keep an open mind to models that broaden our understanding of cosmology beyond $\Lambda$CDM while at the same time remain grounded with data.

A compelling way to study DE without committing to a specific gravity theory is by considering parametric extensions of $\Lambda$CDM. In this direction, a constant DE equation of state, $w_\text{DE} = P_\text{DE}/\rho_\text{DE}$, can be considered, leading to a one parameter extension often called the $w$CDM model. In the same vein, dynamical DE (DDE) models can be considered by taking in additional parameters, such as by the Chevalier-Polarski-Linder (CPL) parametrization \cite{Chevallier:2000qy, Linder:2002et} where the DE equation of state is written as $w_\text{DE}(z) = w_0 - w_a z/(z + 1)$ at redshift $z$. The latest supernovae catalog (Pantheon$+$) constrains the $w$CDM DE parameter to $w_\text{DE} = -0.89 \pm 0.13$ and likewise the CPL DE to $w_0 = -1.81_{-0.60}^{+1.71}$ and $w_a = -0.4^{+1.0}_{-1.8}$ \cite{Scolnic:2021amr, Brout:2022vxf}. The latest Planck observations of the CMB also constrain these two models, leading to $w_\text{DE} = -1.028 \pm 0.031$ on $w$CDM and $w_0 = -0.957 \pm 0.080$ and $w_a = -0.29^{+0.32}_{-0.26}$ on CPL. Other approaches take different parametrizations of the DE functions.

In this work, we consider a phenomenological approach to DE that is grounded on the expansion history. This takes a redshift polynomial expansion of the normalized DE density, $X(z)$, and fits it in the range of the observable data \cite{Wang:2001ht, Wang_2004, Wang:2004ru}. In practice, we identify $X(z)$ through the Friedmann constraint,
\begin{equation}
\label{eq:frd_eq_X}
    E(z)^2 = \Omega_{m0} (1+z)^3 + (1-\Omega_{m0})X(z)\,,
\end{equation}
where $E(z)=H(z)/H_0$ is the normalized Hubble function, $H_0$ is the Hubble constant, and $\Omega_{m0}$ is the matter density parameter or rather the matter fraction today. A spatially flat background is assumed. When $X(z) = 1$, $\Lambda$CDM is recovered; on the other hand, $X(z) \neq 1$ for \textit{any} redshift $z$ supports the DDE picture. The earlier results in this direction have shown that DE density slightly increases with the redshift, but remains consistent with a constant $\Lambda$ at $2\sigma$. More recently, this approach was revisited with more precise data \cite{Cardenas:2014jya, Grandon:2021nls, Bernardo:2021cxi} which lead to the DE density decreasing to lower values, even negative ones, at intermediate redshifts, $z \sim 1$. Other studies \cite{Akarsu:2021fol, Akarsu:2019hmw} also concur with negative DE densities at late times and in \cite{Bernardo:2021cxi} it was shown that nonparametric statistical analysis support this.

The DE phenomenological function $X(z)$ influences all aspects of cosmological evolution. For instance the luminosity distance, $d_L(z)$, measuring the brightness of luminous objects in the sky at a certain redshift,
\begin{equation}
\label{eq:luminosity_distance}
    d_L(z) = \frac{c(1+z)}{H_0}\int_0^z \frac{dz'}{E(z')}\,,
\end{equation}
is directly modified by $X(z)$ through changes in the normalized expansion rate $E(z)$. An advantage of working directly with $X(z)$ over the $w$CDM and CPL models, which are both written at the level of the DE equation of state, is preserving better the dynamical information in the background \cite{Bernardo:2021cxi}. For this work, we take in a quadratic and cubic parametrizations of the DE function $X(z)$ given by
\begin{equation}\label{inter1}
    X(z)= 1+ \left(4x_1 -x_2-3\right) \left( \dfrac{z}{z_m} \right) - 2 \left(2 x_1-x_2-1 \right) \left( \frac{z}{z_m} \right)^2\,
\end{equation}
and
\begin{equation}
\label{eq:cubic}
\begin{split}
X(z) = 1
& + \dfrac{1}{2} \left(-11 + 18 x_1 - 9 x_2 + 2 x_3\right) \left( \frac{z}{z_m} \right) \\
& - \dfrac{9}{2} \left(-2 + 5 x_1 - 4 x_2 + x_3 \right) \left( \frac{z}{z_m} \right)^2 + \dfrac{9}{2} \left(-1 + 3 x_1 - 3 x_2 + x_3 \right) \left( \frac{z}{z_m} \right)^3\,,
\end{split}
\end{equation}
respectively, where the $x_i$'s are the DE parameters and $z_m$ is the maximum redshift in the data. In particular, departures of the DE parameters from unity, $x_i \neq 1$ for any $i$, imply a DDE component in the late Universe, while $x_i = 1$ for all $i$ support the canonical cosmology. Recent work has shown support to DDE interpretation \cite{Cardenas:2014jya, Grandon:2021nls, Bernardo:2021cxi}.

{
Our motivation to parametrize the normalized DE density $X(z)$ \cite{Wang:2001ht} is it emulates DDE background cosmological features more directly than the microscropic equation of state $w(z)$, which is for example two integrations away from the SNe observables. The particular parametrizations studied in this work result from an interpolation method based on a Pade--motivated ansatz, $X(z) \sim x_i \prod_i (z - z_i)/(z_0 - z_i)$, on the redshift structure of DE. The Pade--like expansion, first described in \cite{Wang:2001ht} and further tested in light of more recent cosmological data in \cite{Cardenas:2014jya}, enables a very simple interpretation of the free parameters, e.g., the value of the normalized DE density at the maximum and half maximum redshift in the data. Thus, utilizing this methodology makes transparent the comparison of the constraints with earlier works, highlighting the impact that newer, more stringent data sets have on cosmological parameters. Moreover, going up to a cubic parametrization allowed to study and confirm the trends found with less degrees of freedom \cite{Grandon:2021nls, Bernardo:2021cxi}.
}

{
On this note, we add that a power series, $X(z) - 1 = \sum_{i\geq 1} y_i z^i$, would have been equivalent way to (3) and (4), or rather there is a linear map $\{x_i = c_i y_i\}$, e.g., $y_1 = (4 x_1 - x_2 - 3)/z_m$ in (3), such that DE evolution can be alternatively inferred when $y_i \neq 0$. In this work, on the other hand, we proceed with the convention $\{x_i\}$ which was established in earlier works \cite{Cardenas:2014jya, Grandon:2021nls, Bernardo:2021cxi} to make comparison clearer.
}

In what follows, we refer to the quadratic and cubic realizations of $X(z)$ as $X_2$CDM and $X_3$CDM, respectively. We proceed to compare the above DDE models and $\Lambda$CDM by means of the ABC to see whether a constant $\Lambda$ or DDE is a statistically better description of the data. We next describe the data sets we shall consider in the statistical analysis.

\section{Cosmological data}
\label{sec:cosmodat}

\begin{figure}[h!]
    \centering
	\subfigure[]{
		\includegraphics[width = 0.48 \textwidth]{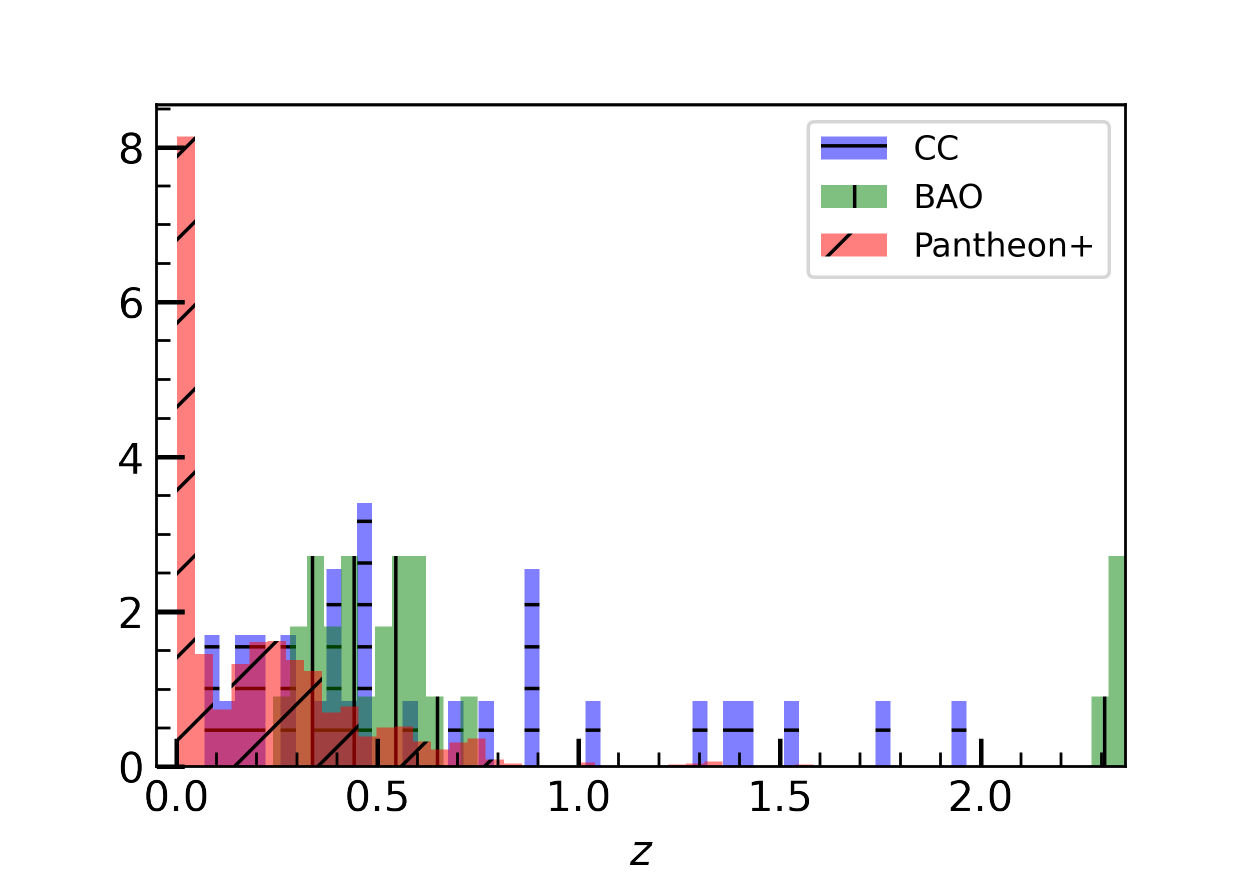}}
    \subfigure[]{
		\includegraphics[width = 0.42 \textwidth]{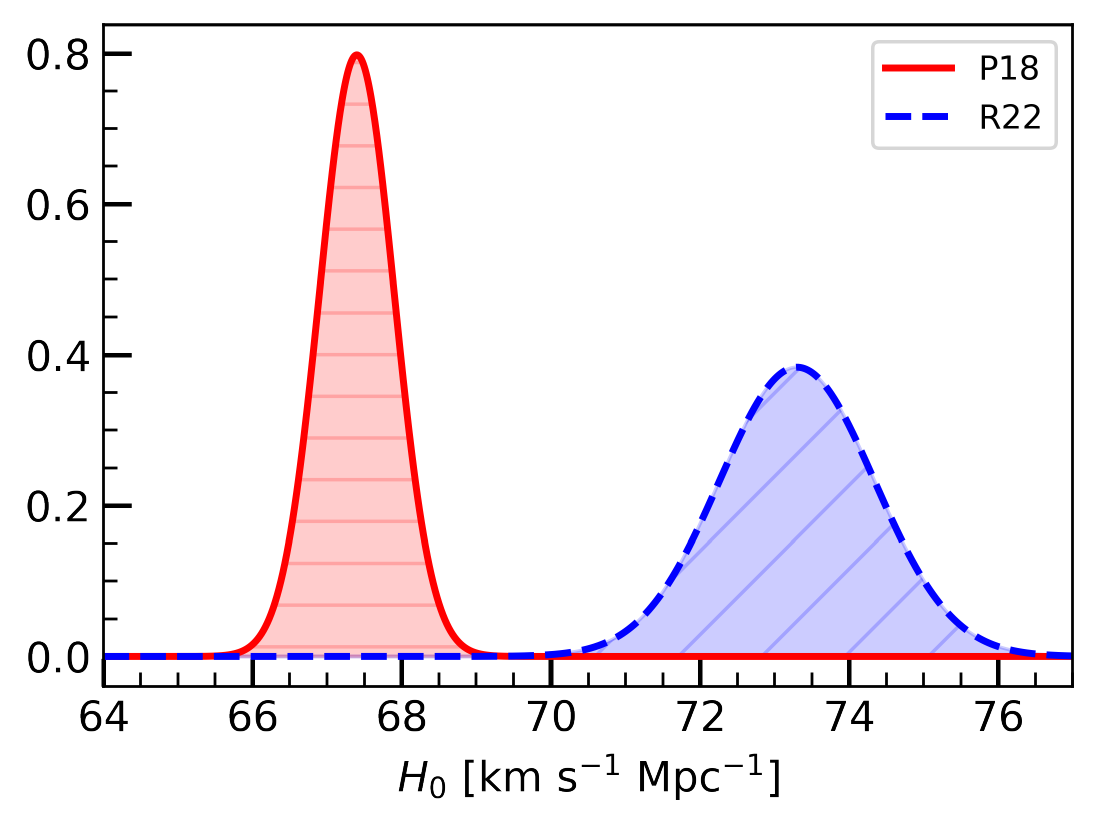}}
    \caption{(a) Redshift distribution of the Hubble data sets (CC, BAO, SNe) used in this work and (b) two Hubble constant measurements, $H_0 = 67.4 \pm 0.5$ km s$^{-1}$Mpc$^{-1}$ \cite{Aghanim:2018eyx} and $H_0^\text{R22} = 73.30 \pm 1.04$ km s$^{-1}$Mpc$^{-1}$ \cite{Riess:2021jrx}, reflecting the present Hubble tension.}
    \label{fig:histdata}
\end{figure}

In this study, we consider the compiled expansion measurements coming from cosmic chronometers (CC), Supernovae type Ia (SNIa) and baryonic acoustic oscillations (BAO). Also, we use latest reporting of the Hubble constant from the Planck collaboration giving $H_0^\text{P18} = 67.4 \pm 0.5$ km s$^{-1}$ Mpc$^{-1}$ \cite{Aghanim:2018eyx}, together with their reporting of the matter and radiation density parameters given respectively as $\Omega_{m0} h^2 = 0.1430 \pm 0.0011$ and $\Omega_{r0} h^2 = 4.15 \times 10^{-5}$. This reported value of the Hubble constant can act as a prior on the Hubble parameter in the ensuing analysis. Moreover, the Planck result is based on the use of observational data of the cosmic microwave background radiation together with $\Lambda$CDM. We also include the latest reported Hubble constant from the SH$0$ES team who give $H_0^\text{R22} = 73.30 \pm 1.04$ km s$^{-1}$ Mpc$^{-1}$ \cite{Riess:2021jrx}. This estimate the of Hubble constant relies on the Hubble flow and is calibrated by Cepheids, and so does not depend on any cosmological model. Figure \ref{fig:histdata} shows the redshift distribution of the CC, SNe, and BAO data sets and the Hubble constant measurements by the Planck and the SH$0$ES missions to be considered in our analysis. Other interesting studies have produced constraints on the Hubble constant but we consider these two since a significant tension appears between their values, and has led to intense study of potential resolutions.

\noindent Below, we briefly describe each data set and how they are considered in our constraint analysis.

\begin{itemize}
    \item Cosmic chronometers (CC) -- We adopt the 31 data points found using the CC technique. This measurement technique allows for Hubble information to be directly derived from observations up to roughly $z\lesssim2$. It is based on the differential aging method as applied to passively-evolving galaxies found to be separated by small redshift intervals, from which one can compute $\Delta z/\Delta t$ directly \cite{Jimenez:2001gg}. In the present study, we utilize data points compiled from Refs. \cite{2014RAA....14.1221Z,Jimenez:2003iv,Moresco:2016mzx,Simon:2004tf,2012JCAP...08..006M,2010JCAP...02..008S,Moresco:2015cya}, which are independent of the Cepheid distance scale, as well as any dependence on cosmology. On the other hand, these points do depend on stellar modeling but this is based on robust stellar population synthesis techniques \cite{Gomez-Valent:2018hwc,Lopez-Corredoira:2017zfl,Verde:2014qea,2012JCAP...08..006M,Moresco:2016mzx}. For the MCMC analysis, the corresponding $\chi^2_{\rm H}$ then turns out to be
    \begin{equation}
        \chi^2_{\rm H}(\Theta)= \sum_{i=1}^{31} \frac{\left(H(z_i,\,\Theta)-H_\mathrm{obs}(z_i)\right)^2}{\sigma_{\rm H}^2(z_i)}\,,
    \end{equation}
    where $H(z_i,\Theta)$ are the theoretical Hubble parameter values at redshift $z_i$ with model parameters $\Theta$, $H_\mathrm{obs}(z_i)$ are the corresponding measured values of the Hubble parameter at $z_i$ with observational error of $\sigma_{\rm H}(z_i)$.
    \item Pantheon+ (SNe) -- The SNe data set consists of 1701 SNIa relative luminosity distance measurements spanning the range $0.01<z<2.3$ \cite{Scolnic:2021amr,Riess:2021jrx,Brout:2021mpj}. The public release of the SN catalog is corrected for systematics, including but not limited to the stretching of the light--curve, the color at maximum brightness and the stellar mass of the host galaxy. These apparent magnitude data points need to be calibrated by an absolute magnitude $M$, and so this will form part of the parameter set for the MCMC analyses, while in other cases we will deal directly with the apparent magnitude. For the MCMC analysis, the connection with the Hubble diagram is made through the distance modulus relation
    \begin{equation}
        \mu(z_i,\,\Theta)=5\log_{10}\left[D_L(z_i,\,\Theta)\right]+M\,,
    \end{equation}
    at redshift $z_i$ via the corresponding computation of the luminosity distance
    \begin{equation}
        D_L(z_i,\Theta)=c\,(1+z_i)\int_0^{z_i}{\frac{\mathrm{d}z'}{H(z',\,\Theta)}}\,,
    \end{equation}
    where $c$ is the speed of light, and the nuisance parameter $M$ encodes the Hubble constant which has to be marginalized over in the MCMC analyses. Hence, the $\chi^2_\mathrm{SN}$ value can be specified by \cite{2011ApJS..192....1C}
    \begin{equation}
        \chi_{\text{SN}}^2(\Theta) = \left(\Delta\mu(z_i,\,\Theta)\right)^{T} C_{\text{SN}}^{-1}\, \Delta\mu(z_i,\,\Theta)+\ln\left({\frac{S}{2\pi}}\right)-\frac{k^2(\Theta)}{S}\,,
    \end{equation}
    where $C_{\text{SN}}^{}$ is the total covariance matrix, $S$ is the sum of all the components of $C_{\rm SN}^{-1}$, while $k$ is given by
    \begin{equation}
        k(\Theta)={\left(\Delta\mu(z_i,\,\Theta)\right)^{T}\cdotp C_{\text{SN}}^{-1}}\,,
    \end{equation}
    with $\Delta\mu(z_i,\,\Theta)=\mu(z_i,\,\Theta)-\mu_{\text{obs}}(z_i)$.
    
    {
    We note that with Pantheon$+$ data, it is no longer necessary to marginalize over $M$ since the distance--modulus $\mu = m - M$ for each SNIa are given relative to their Cepheid hosts. On the other hand, we use the provided SNIa apparent magnitudes, $m$, which remain available in Pantheon$+$ and are conservative to the calibration of the distance--ladder, and so can be analyzed similarly with the Pantheon 2018 data. Otherwise, using $\mu$ in Pantheon$+$ together with $H_0$ priors would be tantamount to marginalizing twice over $M$, because $M$ and $H_0$ are degenerate in SNIa cosmology.
    }
    \item Baryonic Acoustic Oscillations (BAO) -- Here, we use the 26 data points coming from line--of--sight BAO measurements \cite{2012MNRAS.425..405B, Chuang:2013hya, BOSS:2013igd, BOSS:2014hwf, Bautista:2017zgn}, which are not correlated and so do not require a covariance matrix. BAO occur as perturbations in the early Universe baryon-photon plasma during the drag epoch which then become frozen in the large scale structure that evolves over time. This mechanism produces a standard ruler by which to infer cosmic expansion. In particular, the BAO provides the expansion rate at a redshift $z_i$ through the combination $H(z_i) r_d$ where $r_d = 147.74$ Mpc is the sound horizon radius during the drag epoch in $\Lambda$CDM. 
    
    {
    Altogether, the CC and BAO expansion rate measurements we considered for this work is summarized in Table A.3 of \cite{Bernardo:2021cxi}.
    }
\end{itemize}


\section{ABC and MCMC}
\label{sec:abcvsmcmc}

We review the traditional cosmological parameter inference algorithm based on the Markov chain Monte Carlo (MCMC) algorithm and briefly introduce the ABC for model selection and parameter estimation.

\subsection{Markov chain Monte Carlo}
\label{subse:mcmc}

The traditional parameter estimation in cosmology is based on the Bayes theorem, that is, in a model,
\begin{equation}
\label{eq:bayestheorem}
    P\left( \theta \vert D \right) = \dfrac{\mathcal{L}
(D\vert \theta)P(\theta)}{P(D)} \,,
\end{equation}
where $P(\theta\vert D)$ is the posterior for the parameter(s) $\theta$ given the data $D$, $\mathcal{L}$ is the likelihood, and $P(\theta)$ is the prior distribution on the parameters. Through \eqref{eq:bayestheorem}, the posterior is inferred through a random series of steps in the parameter space $\theta$ such that $P\left(\theta\vert D\right) \propto P\left(D\vert \theta\right) P\left(\theta\right)$. One way to realize this is by means of the Metropolis-Hashtings algorithm, where basically the posterior is sequentially approached by compelling the random walker to step more frequently in the direction of increasing likelihood. In this way, the parameters are estimated by where it spends its most time (the mean) and how far it deviates from this place (the error) in the parameter space. In general, the shape of the posterior is estimated to desired precision by letting the walker make a sufficient number of steps.

Model selection is also tackled in this Bayesian setting by calculating the evidence for each model individually, and then comparing the results after \cite{Trotta:2008qt}. Compared with mere parameter estimation, Bayesian model selection demands more computationally and relies on nested sampling for the calculation of the evidence. Nonetheless, when the posteriors are approximately Gaussian, as is often the case in cosmology, several other assessment tools such as the information criteria can stand as a reliable surrogate to the evidence \cite{Trotta:2008qt}. These can be obtained quite straightforwardly as soon as the parameters are estimated per model. In this way, it can be revealed which model is preferred or ruled out by the data by which has the lower or larger values of the information criteria.

In this work we consider flat priors on all the dynamical dark energy parameters, and a Gaussian prior on $\Omega_{m0}h^2$ derived from Planck satellite 2018 cosmological constraints \cite{Aghanim:2018eyx}. We also consider two different Gaussian priors on $H_0$, as presented in Section \ref{sec:cosmodat}. We perform the MCMC analysis using the public code emcee \cite{emcee}.

\subsection{Approximate Bayesian computation}
\label{sec:abcsmc}

ABC is a biology inspired inference algorithm that is the most useful whenever the likelihoods are unknown, or intractable. As it is, this is often the case when it comes to model selection in cosmology, such that different models enter the picture, and the job of the cosmologist is to single out which one best describes the data. However, the likelihood across different cosmological models is as difficult to grasp, as it is with the parameter within the models. The ABC instead constructs a joint model space, where each model is assigned a probability, say, model $1$ has $30\%$, model $2$ has $15\%$, and so on. The sequential Monte Carlo spin on the ABC lets the model posterior evolve, in such a way that after each generation, the models and the parameters within become better approximations to the data, until only one model remains. This makes the resulting ABC-SMC algorithm as a natural model selection tool. The ABC has been applied successfully in biology and chemistry, as well as in astrophysics and cosmology \cite{2009arXiv0911.1705T, Akeret:2015uha, 2013ApJ...764..116W, 10.3389/fbuil.2017.00052, 2017A&C....19...16J, Ishida:2015wla}. We delve deeper into the overall details of the algorithm below.

ABC, as a model selection algorithm, aims to approximate the posterior distribution 
\begin{equation}
    P \left( m | D \right) \propto L \left( D | m \right)\,,
\end{equation}
where $m$ are the models and $L$ is the likelihood between a model $m$ and the data $D$. The likelihood across models is however often unknown. ABC proceeds to do so by bypassing this likelihood step, and sequentially getting to the posterior efficiently by relying on a distance function $\Delta(R)$, which measures how far a prediction $D^*$, drawn from a model $m^*$ with parameters $\theta^*$, is away from the data. Typical choices of the distance function for noise free data are the absolute distance and the mean-squared error. In physics, where measurement uncertainties are given utmost interest, a reasonable choice is the chi squared, estimating how many standard errors away a prediction is from the mean of the observation. A drawn `particle' $\left( m^*, \theta^* \right)$ is accepted if its distance is smaller than the tolerance $\epsilon$, in symbols, $\Delta(D - D^*) \leq \epsilon$. The tolerance is sequentially tightened with each `generation', leading to a more stringent constraint on the model space and the parameters. One ingredient which makes the ABC powered by the sequential Monte Carlo approach numerically efficient is that the samples of the newer populations are drawn from the previous one. In other words, every new versions of a population come out better than their predecessors. In model selection and parameter estimation, this translates to inevitably ending up with a better, more reliable, approximation of both the model and the parameter posterior distributions.

{
Another point to be aware of with ABC is that since a likelihood does not exist, as in MCMC, it is not a trivial task to compare the outcomes of both approaches in terms of the parameter constraints that they arrive at. We stress nonetheless that the ABC-SMC scheme performs model selection and parameter estimation \textit{at the same time} and so the outputs necessarily hinge on which models participate in the competition and how long the natural selection take place \cite{Toni_2008}. For one, this implies that the errors estimated by the ABC and MCMC have different meanings, and so we should not overthink their comparison. Another way to view this is that MCMC is but one generation of ABC, which reflect on ABC estimated posteriors being narrower compared with their MCMC counterparts. We lastly point out that the ABC-SMC algorithm also implicitly penalizes models with larger number of parameters by lowering their particles' acceptance chances \cite{Toni_2008}.
}

For more details\footnote{
{
The ABC-SMC algorithm in greater detail \cite{Toni_2008, 2009arXiv0911.1705T, 2009arXiv0910.4472T}:
\begin{enumerate}[1]
    \item Initialize tolerances $\{\epsilon_t\}$ such that $\epsilon_{t + 1} < \epsilon_{t}$ and set population indicator $t = 0$;
    \item \begin{enumerate}
        \item[0] Set particle indicator $i = 1$;
        \item[1] Sample $m^*$ from $\pi(m)$;
        \begin{itemize}
            \item[] If $t = 0$, sample $\theta^{**}$ from $\pi\left(\theta\left(m^*\right)\right)$.
            \item[] If $t > 0$, sample $\theta^*$ from previous population $\{ \theta\left(m^*\right)_{t-1} \}$ with weights $w\left(m^*\right)_{t-1}$.
            \item[] Perturb the particle $\theta^*$ to obtain $\theta^{**} \sim K_t\left(\theta|\theta^*\right)$.
            \item[] If $\pi\left(\theta^{**}\right) = 0$, return to 2.1.
            \item[] Simulate a candidate data set $x^* \sim f\left(x|\theta^{**},m^{**}\right)$.
            \item[] If $d\left(x^*, x_0\right) \geq \epsilon_t$, return to 2.1.
        \end{itemize}
        \item[2] Set $m_t^{i} = m^*$ and add $\theta^{**}$ to the population of particles $\{ \theta\left(m^*\right)_t \}$, and calculate its weight as
        \begin{equation} \nonumber
            w_t^{(i)} =
            \begin{cases}
            1 & , {\rm if} \ t = 0 \\
            \dfrac{\pi\left( \theta^{**} \right)}{\sum_{j=1}^N w_{t-1}^{(j)} K_t\left( \theta_{t-1}^{j}, \theta^{**} \right) } & , {\rm otherwise} \,.
            \end{cases}
        \end{equation}
        \begin{itemize}
            \item[] If $i < N$ set $i = i + 1$, go to 2.1. 
        \end{itemize}
    \end{enumerate}
    \item For every $m$, normalize the weights;
    \begin{itemize}
        \item[] For every $t < T$, set $t = t + 1$, go to 2.0.
    \end{itemize}
\end{enumerate}
}
}, we refer the reader to \cite{Toni_2008, 2009arXiv0911.1705T, 2009arXiv0910.4472T} for an outstanding introduction to ABC. We also refer to python package `pyabc' \cite{10.1093/bioinformatics/bty361}, a one stop implementation of the ABC algorithm with various scientific examples.

In our application, we use the ABC to compare between the $\Lambda$CDM model and the DDE models, $X_2$CDM and $X_3$CDM. We consider the chi squared as a distance function, and employ pyabc for implementing the ABC with the sequential Monte Carlo approach. We take the CMB constraint on $\Omega_{m0}h^2$ \cite{Aghanim:2018eyx}, the combination of the matter density and Hubble constant, determined by the peak of CMB damping tail, as a prior for DDE. The dark energy parameters take on flat priors, while the Hubble constant takes flat or Gaussian priors based on the estimates by the Planck \cite{Aghanim:2018eyx} and the SH$0$ES \cite{Riess:2021jrx} collaborations.

\section{Results}
\label{sec:results}

We present the main results of our paper, that is, our constraints on DDE obtained through ABC and MCMC methods. We divide the discussion in three parts depending on whether we consider the BAO data set and Hubble constant priors.

\subsection{SNe \texorpdfstring{$+$}{} CC}
\label{subsec:snc}

We begin with our most conservative data set, that is, of the SNe and CC, both of which can be studied independent of the context of a cosmological model. The resulting ABC run is shown in the following plots.

\begin{figure}[h!]
    \centering
	\subfigure[ \ First thirty generations ]{
		\includegraphics[width = 0.47 \textwidth]{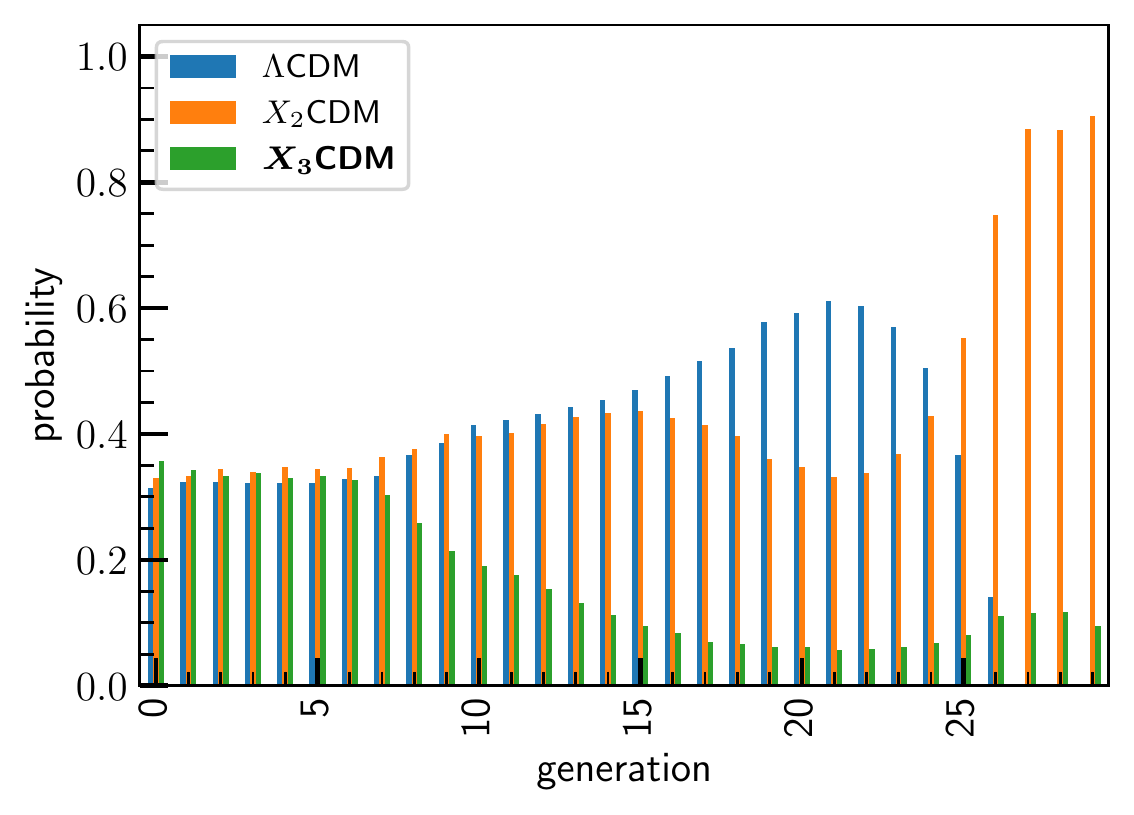}
		}
	\subfigure[ \ After forty five generations ]{
		\includegraphics[width = 0.47 \textwidth]{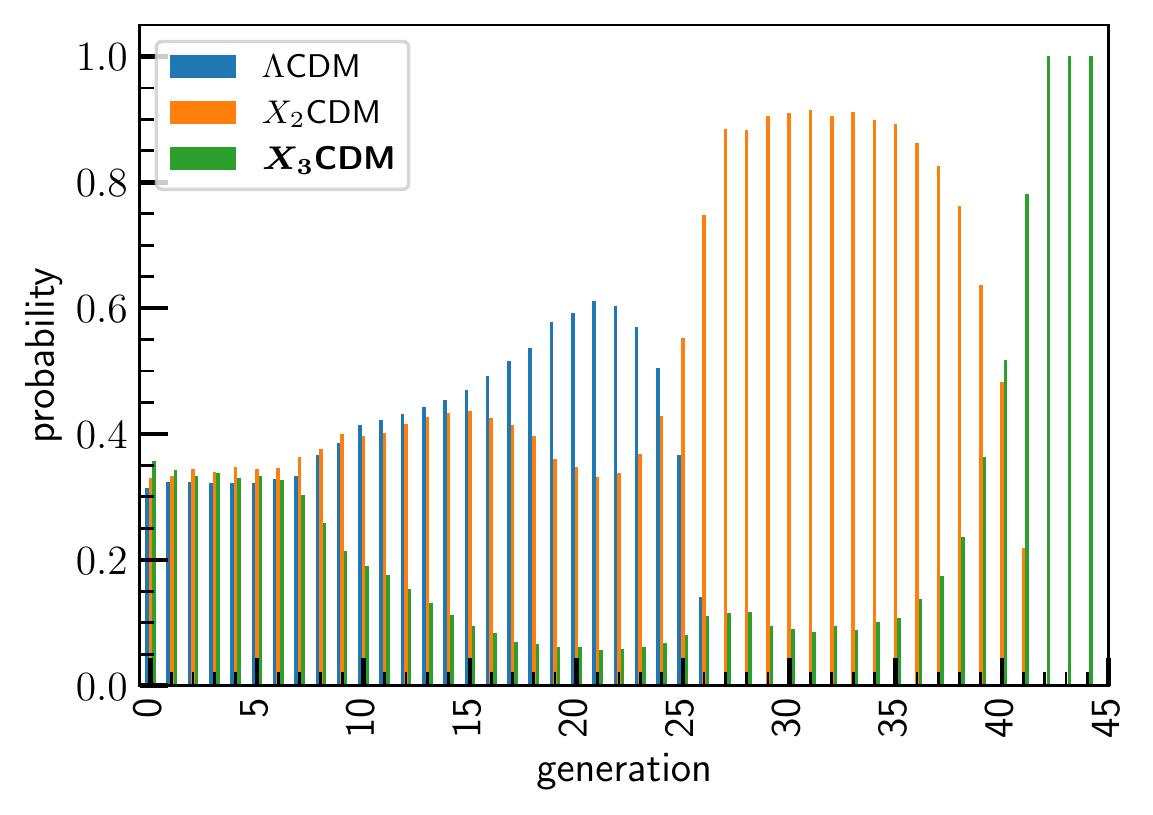}
		}
    \caption{Model posterior evolution constrained by the SNe and CC data sets.}
    \label{fig:modelX23post_noBAO}
\end{figure}

Figure \ref{fig:modelX23post_noBAO} shows the evolution of the model posterior, illustrating what ABC does naturally.

This reveals a model space that contains $\Lambda$CDM as well as both DDE models, $X_2$CDM and $X_3$CDM. As alluded to in the previous section, the likelihood that describes any model space is unknown, and ABC stands out in this regard as a likelihood free inferencing method. We find here that after the calibration sample, that is, the first generation, when each model is given roughly the same weight, the natural evolution proceeds with each generation passing. In particular, in Figure \ref{fig:modelX23post_noBAO}, each model have roughly the same weight after $7$ generations. After this, the probabilities of both $\Lambda$CDM and $X_2$CDM have increased, consequently decreasing that of $X_3$CDM. However, after some more generations, the probability of $\Lambda$CDM have decreased with respect to both DDE models. This continued, with its population declining rapidly, until its extinction at generation $27$. At this point, it is up to the DDE models to compete which represents the data better, albeit seemingly, it looks for a while there, some ten generations, that $X_2$CDM will be dominating. But then, $X_3$CDM's population started gaining some quality, that it eventually overcome the odds at generation $40$. At generation $42$, it simply came out on top, dominating over $\Lambda$CDM and $X_2$CDM during the process. This demonstrates how ABC works to resolve tensions: one model comes out on top, that is, by natural selection. 

\begin{figure}[h!]
    \centering
	\subfigure[ \ Hubble constant ]{
		\includegraphics[width = 0.47 \textwidth]{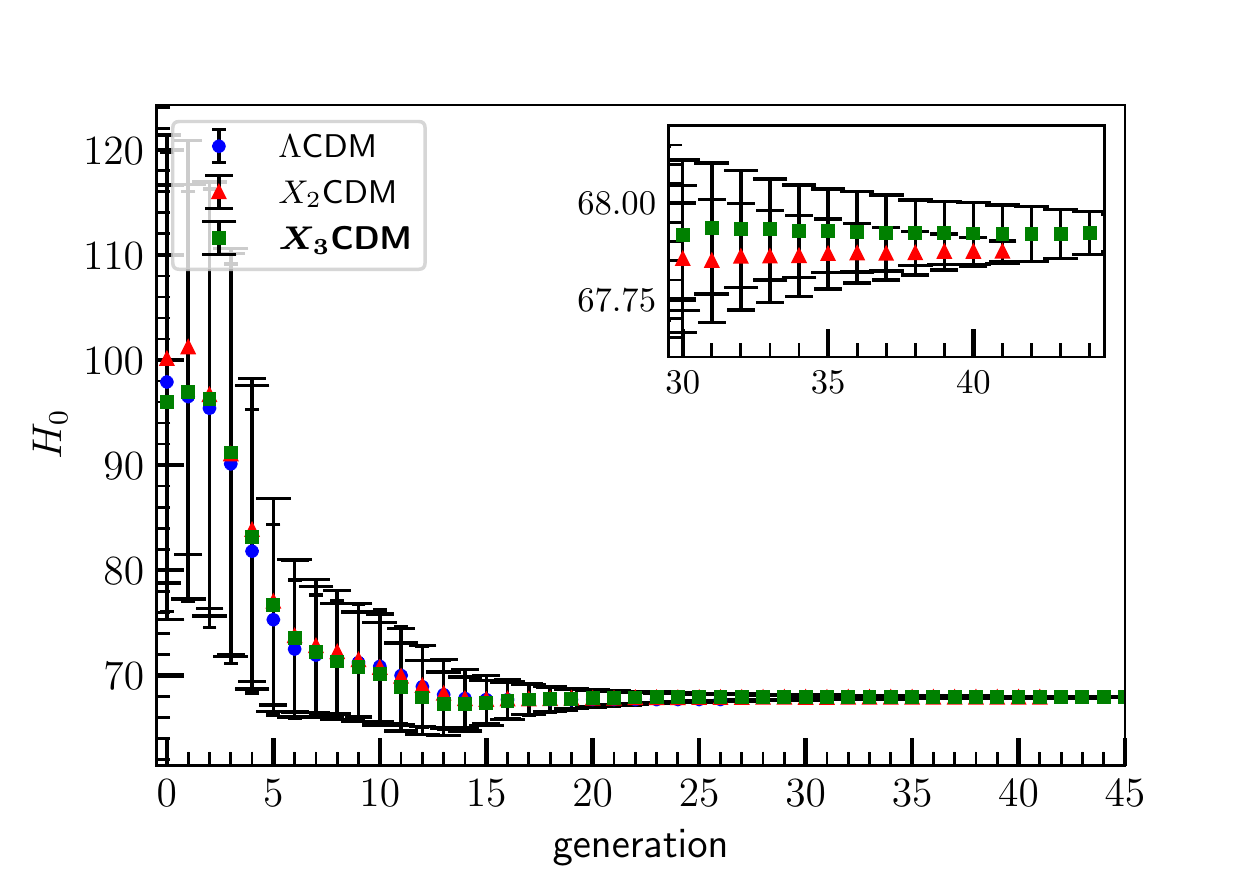}
		}
	\subfigure[ \ matter density ]{
		\includegraphics[width = 0.47 \textwidth]{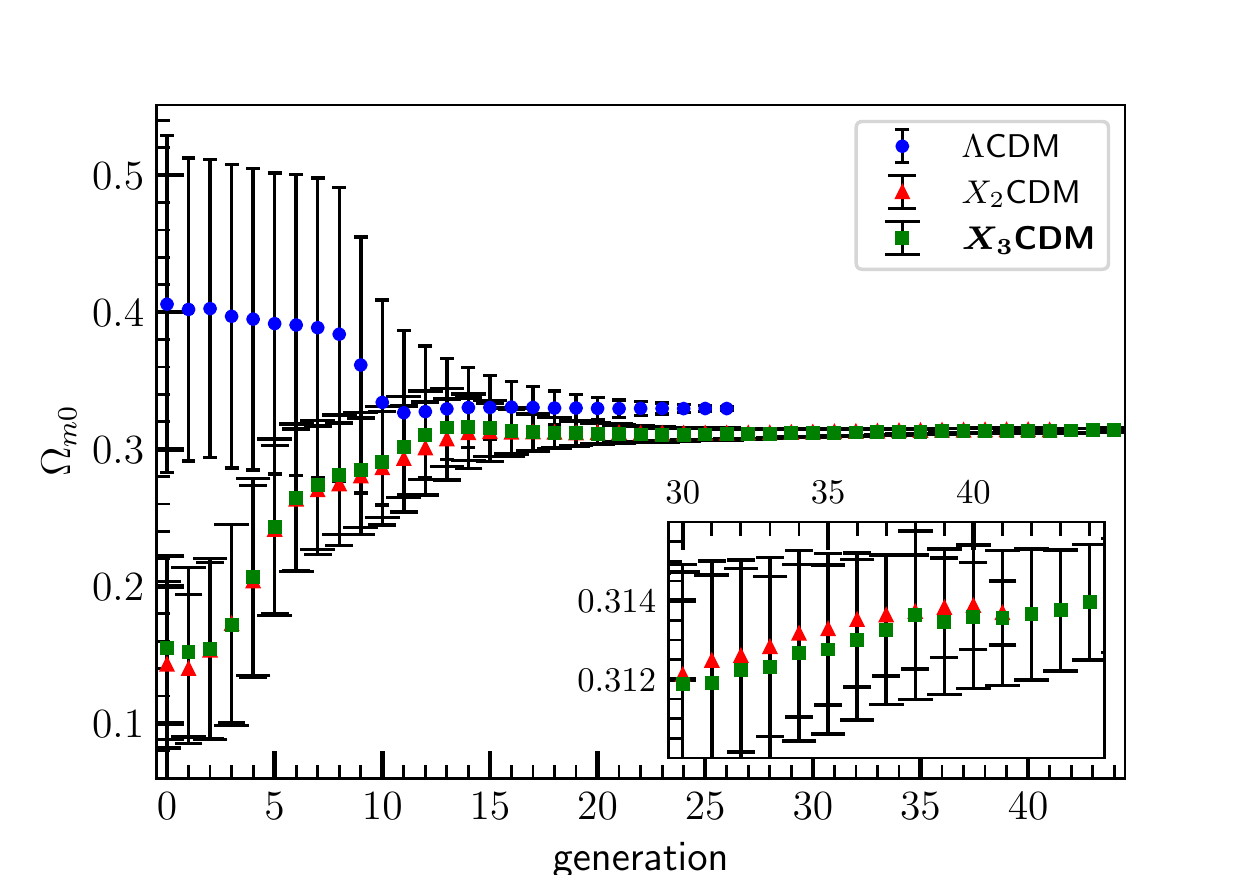}
		}
    \caption{Generational evolution of the model parameters constrained by the SNe and CC data sets. Insets show a zoomed-in view starting from generation thirty.}
    \label{fig:ABCX23H0om0generational_noBAO}
\end{figure}

In addition, ABC estimates the parameters of the models, even the ones that go extinct during the process. Understandably, it is only the parameters of the winning model that makes sense or are meaningful ABC-wise, but for the sake of the discussion we also take the parameters of the models that gone out during the evolution. Figure \ref{fig:ABCX23H0om0generational_noBAO} shows this for the Hubble constant and the matter density during the course of $45$ generations.

Once again, recalling that $\Lambda$CDM gets extinct at generation $27$ while $X_2$CDM has gone out at generation $42$, we present only their parameters for the discussion. As far as the ABC is concerned, only the parameters of the $X_3$CDM model makes sense, as it is the model that survived the natural evolution. It is in this way that ABC overcomes any possible tension within the model space. This is given more light in Figure \ref{fig:ABCX23H0om0generational_noBAO}, where it can be seen that the parameters in all three models are measured in the first few generations, when all three models are still quite competitive. Then, however, at some point, the parameters of the weaker models -- in the context of ABC -- would die out as their populations shrink, leading to the models' extinction. This happened to $\Lambda$CDM at generation $27$ and the $X_2$CDM at generation $42$. In the insets of Figure \ref{fig:ABCX23H0om0generational_noBAO}, it can be seen that after 42 generations, only the parameters of $X_3$CDM continue to be measured.

We may mention as well that the model probability for each generation in Figure \ref{fig:modelX23post_noBAO} can be associated with the model parameters in Figure \ref{fig:ABCX23H0om0generational_noBAO}. It can be seen for instance that at generation zero, the error bars of the Hubble constant are quite too large, and then the estimates only become better with each passing generation. One way to see how this works is to imagine this model space, with $\Lambda$CDM and the DDE models $X_2$CDM and $X_3$CDM, where in each generation, each member of the population of each model passes on a measurement of the model parameters and are evaluated according to their distance from the observation. The population which measures the overall shortest distance to the data receives the edge in the next generation. However, the criterion for an acceptable distance measurement becomes more stringent with age, consequently reducing the population size of the weaker models while adding strength to the better ones. This repeats until only one kind of the population, or rather a model, remains.

We present the ABC statistics of each parameter, together with the corresponding MCMC counterparts, of the resulting ABC run in Table \ref{tab:ABCX23bestfits_noBAO}.

\begin{table}[h!]
    \centering
    \caption{ABC and MCMC parameter statistics of each model constrained with the SNe and CC data sets. For $\Lambda$CDM and $X_2$CDM the parameters are determined by the last surviving populations.}
    \begin{tabular}{|c|c|c|c|c|c|c|} \hline
    \phantom{gg} algo \phantom{gg} & \phantom{gg} model \phantom{gg} & $H_0$ [km s$^{-1}$Mpc$^{-1}$] & \phantom{gggg} $\Omega_{m0}$ \phantom{gggg} & \phantom{gggg} $x_1$ \phantom{gggg} & \phantom{gggg} $x_2$ \phantom{gggg} & \phantom{gggg} $x_3$ \phantom{gggg} \\ \hline \hline
    \multirow{3}{*}{ABC} &
    $\Lambda$CDM & $67.8 \pm 0.1$ & $0.330 \pm 0.001$ & $-$ & $-$ & $-$ \\ 
    & $X_2$CDM & $67.87 \pm 0.03$ & $0.314 \pm 0.001$ & $1.08 \pm 0.01$ & $0.97 \pm 0.02$ & $-$ \\ 
    & \textbf{$\mathbf{X_3}$CDM} & $\mathbf{67.92 \pm 0.03}$ & $\mathbf{0.315 \pm 0.001}$ & $\mathbf{1.07 \pm 0.01}$ & $\mathbf{0.97 \pm 0.03}$ & $\mathbf{0.46 \pm 0.07}$ \\ \hline \hline
    \multirow{3}{*}{MCMC} &
    $\Lambda$CDM & $67.4 \pm 1.6$ & $0.326 \pm 0.017$ & $-$ & $-$ & $-$ \\ 
    & $X_2$CDM & $67.7 \pm 1.8$ & $0.317 \pm 0.025$ & $ 0.97 \pm 0.25$ & $0.5 \pm 1.0$ & $-$ \\ 
    & $X_3$CDM & $67.7 \pm 1.8$ & $0.315 \pm 0.028$ & $1.1 \pm 0.2$ & $0.89 \pm 0.55$ & $0.2 \pm 1.9$ \\ \hline
    \end{tabular}
    \label{tab:ABCX23bestfits_noBAO}
\end{table}

We emphasize that the parameters of the $\Lambda$CDM and $X_2$CDM models are only presented for the discussion, as their populations shrinked to extinction in the history of the ABC. As the result shows, ABC prefers the cubic DDE model, $X_3$CDM, which brings in the cosmological parameter estimates appearing in the bold fonts in Table \ref{tab:ABCX23bestfits_noBAO}. It is interesting to note that the estimate of $X_3$CDM nearly coincides with the last $X_2$CDM generation except that it has one more DE parameter $x_3$. Be that as it may, the ABC natural evolution algorithm lead to DDE, ruling out $\Lambda$CDM in the process.

It is worth commenting on the parameter estimates of the last generations of $\Lambda$CDM and $X_2$CDM. What is interesting here to note is that both Hubble constants emerging from this are consistent with the Planck estimate. In fact, all three Hubble constants from each model can be found inside the $95\%$ confidence intervals of the Planck measurement. We stress however that this does not necessarily imply consistency with the Planck values. The $\Lambda$CDM matter density estimate for instance is quite large, and outside, the Planck estimate \cite{Aghanim:2018eyx}. On the other hand, it is rather intriguing the DDE models turns out to agree with Planck on both the Hubble constant and the matter density, despite the case that the Planck cosmological parameters rely on the $\Lambda$CDM model throughout. We understand this is because of the additional use of the CMB constraint on the combination $\Omega_{m0} h^2$ as a prior in the DDE models \cite{Aghanim:2018eyx, Bernardo:2021cxi}. Since the $H_0$ values resulting from the ABC become consistent with Planck, this constraint compels the matter density to also agree with the Planck data.

\begin{figure}[h!]
    \centering
	\subfigure[ \ Hubble expansion rate ]{
		\includegraphics[width = 0.47 \textwidth]{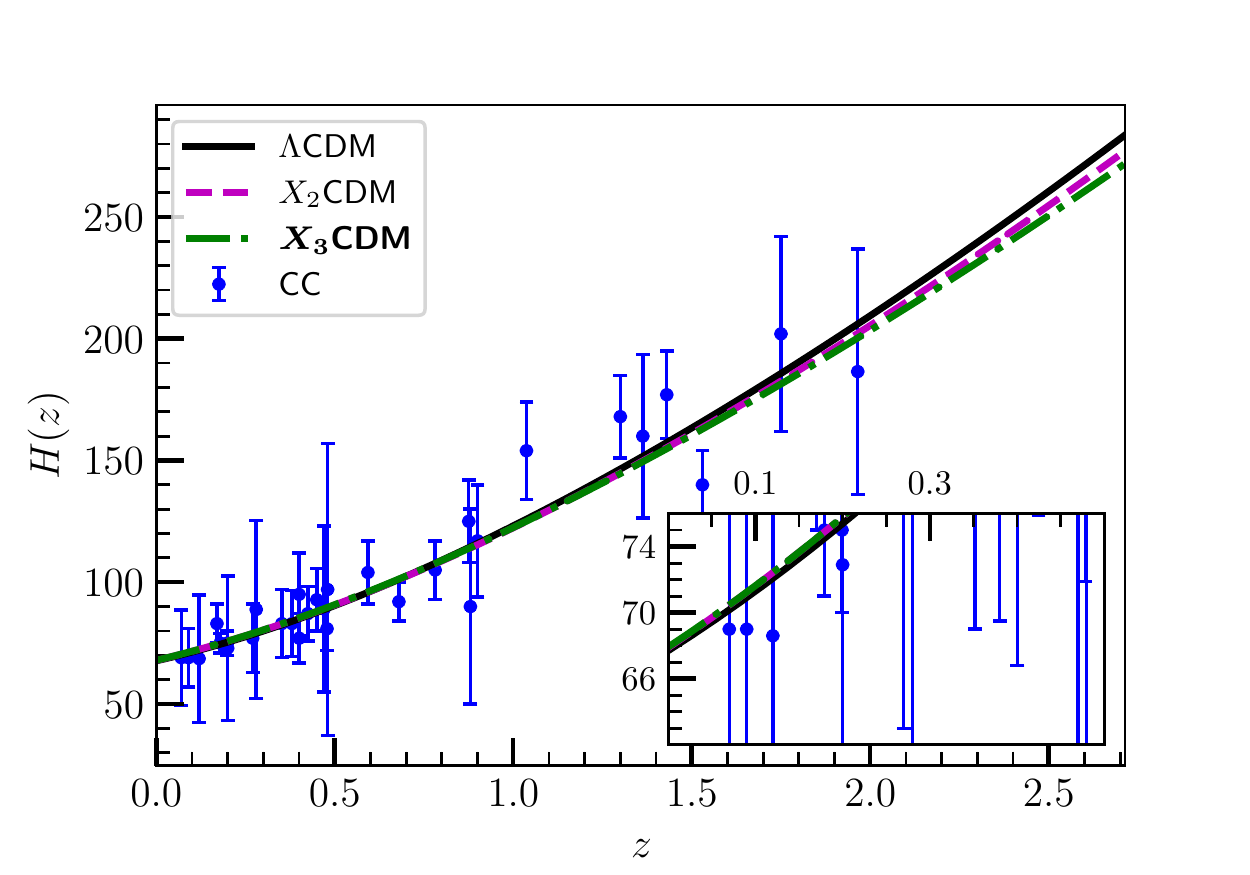}
		}
	\subfigure[ \ SNe brightness ]{
		\includegraphics[width = 0.47 \textwidth]{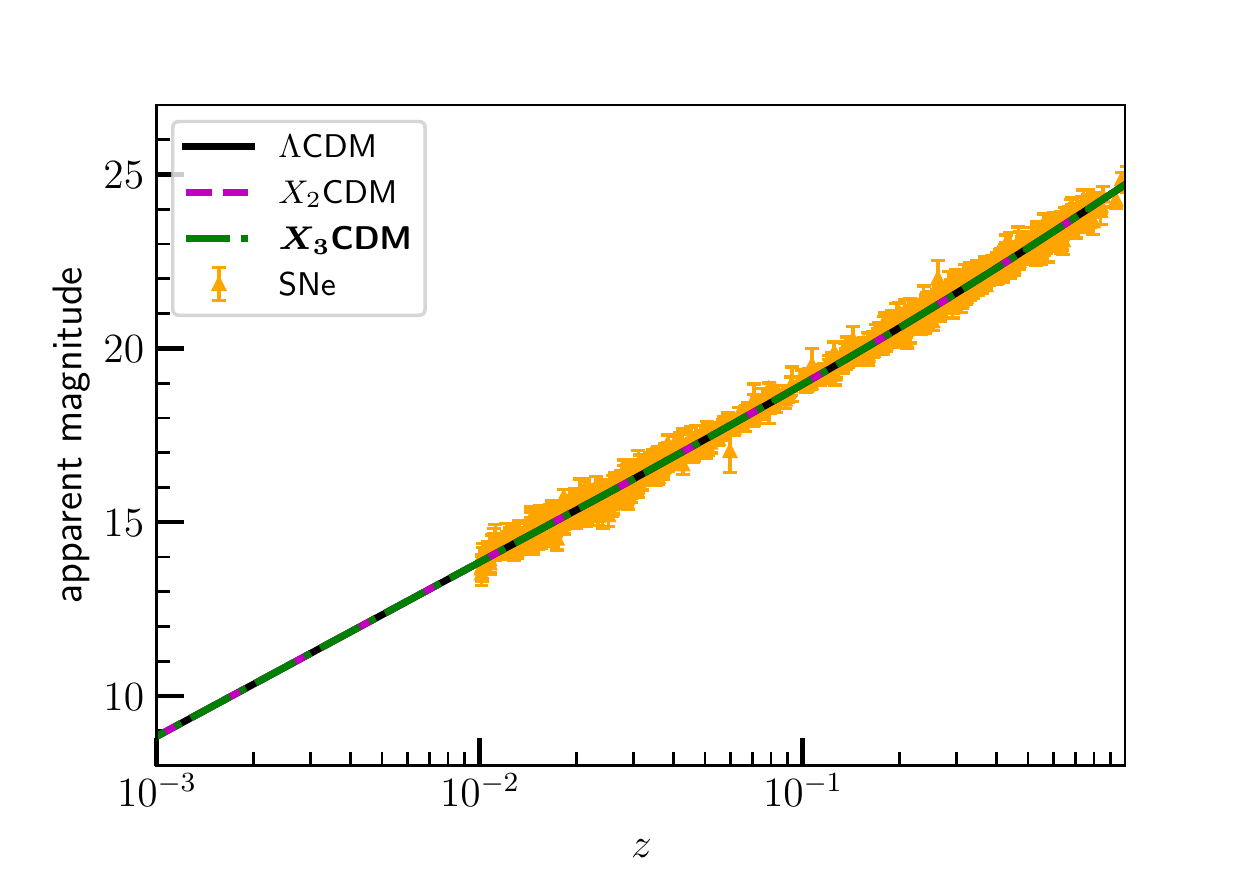}
		}
    \caption{Best fit curves for each model constrained by the SNe and CC data sets after forty five generations. For $\Lambda$CDM and $X_2$CDM, the curves are determined by the last surviving population.}
    \label{fig:ABCbestcurves_noBAO}
\end{figure}

We show in Figure \ref{fig:ABCbestcurves_noBAO} the best fit curves from each model as determined by ABC to allow us to visualize how the algorithm may have selected its overall victor.

We find here that the best fit curves for the SNe magnitudes appear visually identical, understandably because of the density of the data secured in the Pantheon$+$ sample. The noise all in all of the SNe data thus apparently secure the best fit curves to be indistinguishable as shown. {For the Hubble diagram, the CC data is also added}. To distinguish between the models, we resort to the Hubble expansion rate best fit predictions. In this case, we find the three distinct curves, which reflect the consistency with their Hubble constant estimates (Table \ref{tab:ABCX23bestfits_noBAO}) as revealed further in the inset. The high redshift Hubble expansion rate data, $z = O(1)$, appear to play the major role in distinguishing between the models. The $\Lambda$CDM best fit curve for instance turned out to has the highest expansion rate at high redshifts, while the $X_2$CDM best fit runs below this. Instead, the ABC preferred $X_3$CDM has its high redshift Hubble expansion rate below the $\Lambda$CDM and $X_2$CDM best fit curves.

We find this high redshift Hubble expansion to be ABC's characteristic feature that differentiated the models. Thus it makes sense to consider independent measurements of the cosmological expansion in the data.

\subsection{SNe \texorpdfstring{$+$}{} CC \texorpdfstring{$+$}{} BAO}
\label{subsec:sncb}

We present in this section the same analysis as in the previous one, except with a more stringent data set that includes BAO. Since the BAO have tighter measurements of the expansion rate at various redshifts, it is interesting whether this may impact the results.

\begin{figure}[h!]
    \centering
	\subfigure[ \ First twenty generations ]{
		\includegraphics[width = 0.47 \textwidth]{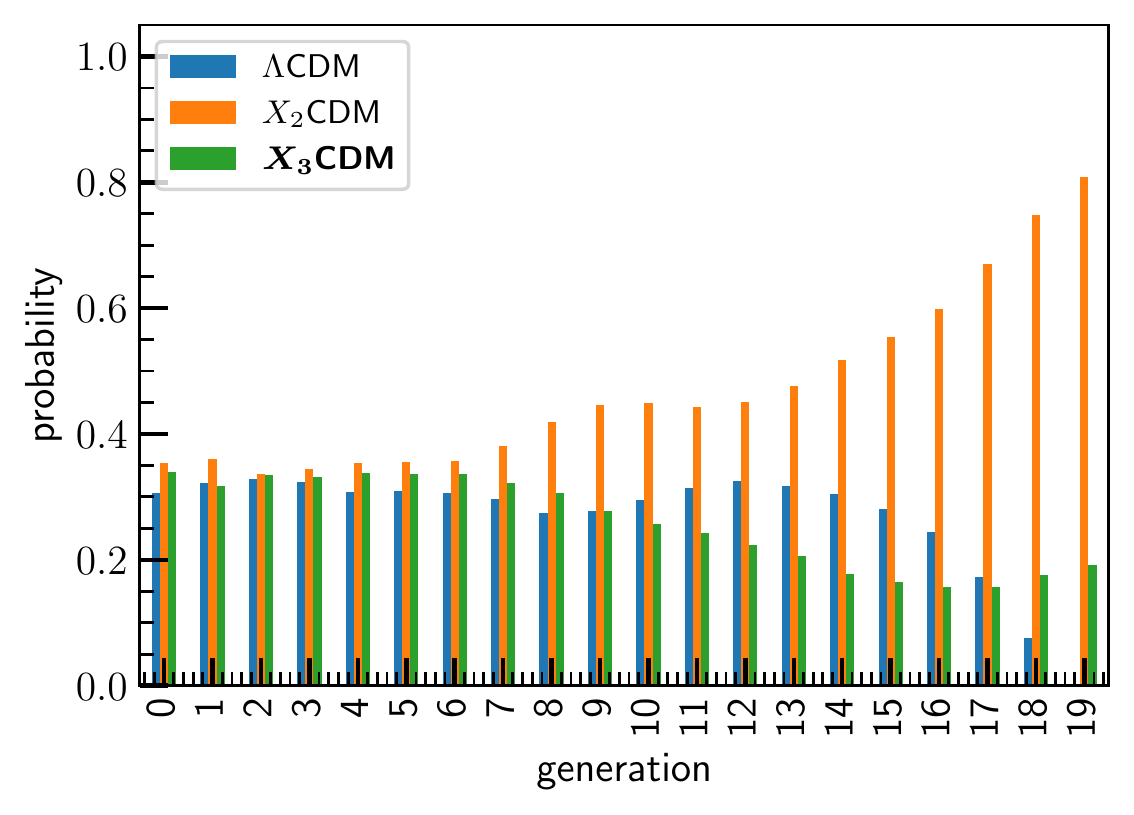}
		}
	\subfigure[ \ After thirty generations ]{
		\includegraphics[width = 0.47 \textwidth]{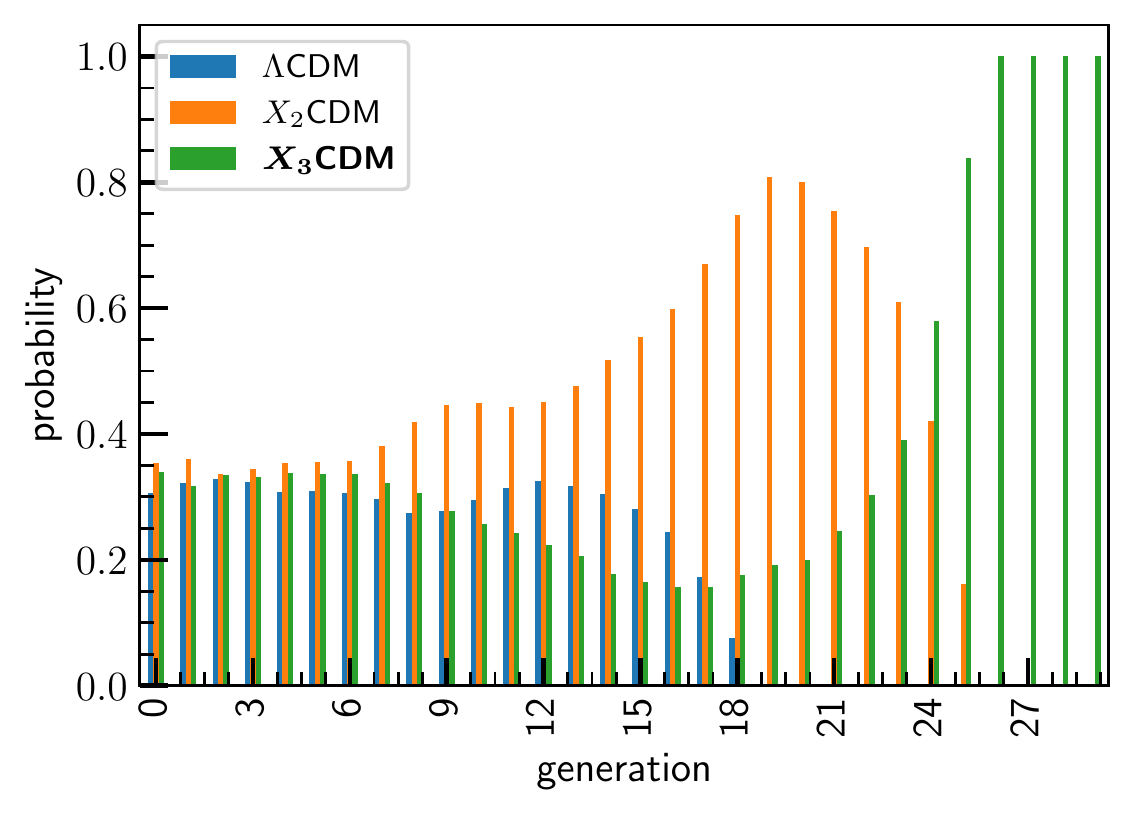}
		}
    \caption{Model posterior evolution constrained by the SNe, CC, and BAO data sets.}
    \label{fig:modelX23post}
\end{figure}

Figure \ref{fig:modelX23post} shows the model posterior evolution with the SNe, CC, and BAO compiled data sets, this time taking about $30$ generations to host a winning model.

Perhaps the most interesting feature that is captured with this analysis is that it tells nearly the same story, as far as the evolution goes, as without the BAO. That is, it starts with the calibration when all the models have roughly the same probabilities for about $5-6$ generations. Then, beginning at generation $7$ in particular, we see that $X_2$CDM becomes noticeably more probable compared with the other two models. $X_2$CDM would after this continue to increase its probability, while that of $\Lambda$CDM would shrink, until eventually its population collapses at generation $19$, leaving the DDE models to compete for survival. At this point, $X_2$CDM holds the edge, being quite the more probable DDE model, compared with $X_3$CDM. This competition between the two DDE populations would not take too long, with $X_2$CDM seemingly being the favorable one for a short time. The $X_3$CDM population starts to defy the odds and dominate eventually. At generation $24$, the $X_3$CDM model takes over, surviving the ABC natural selection, whereas the $X_2$CDM model eventually becomes extinct at generation $26$.

\begin{figure}[h!]
    \centering
	\subfigure[ \ Hubble constant ]{
		\includegraphics[width = 0.47 \textwidth]{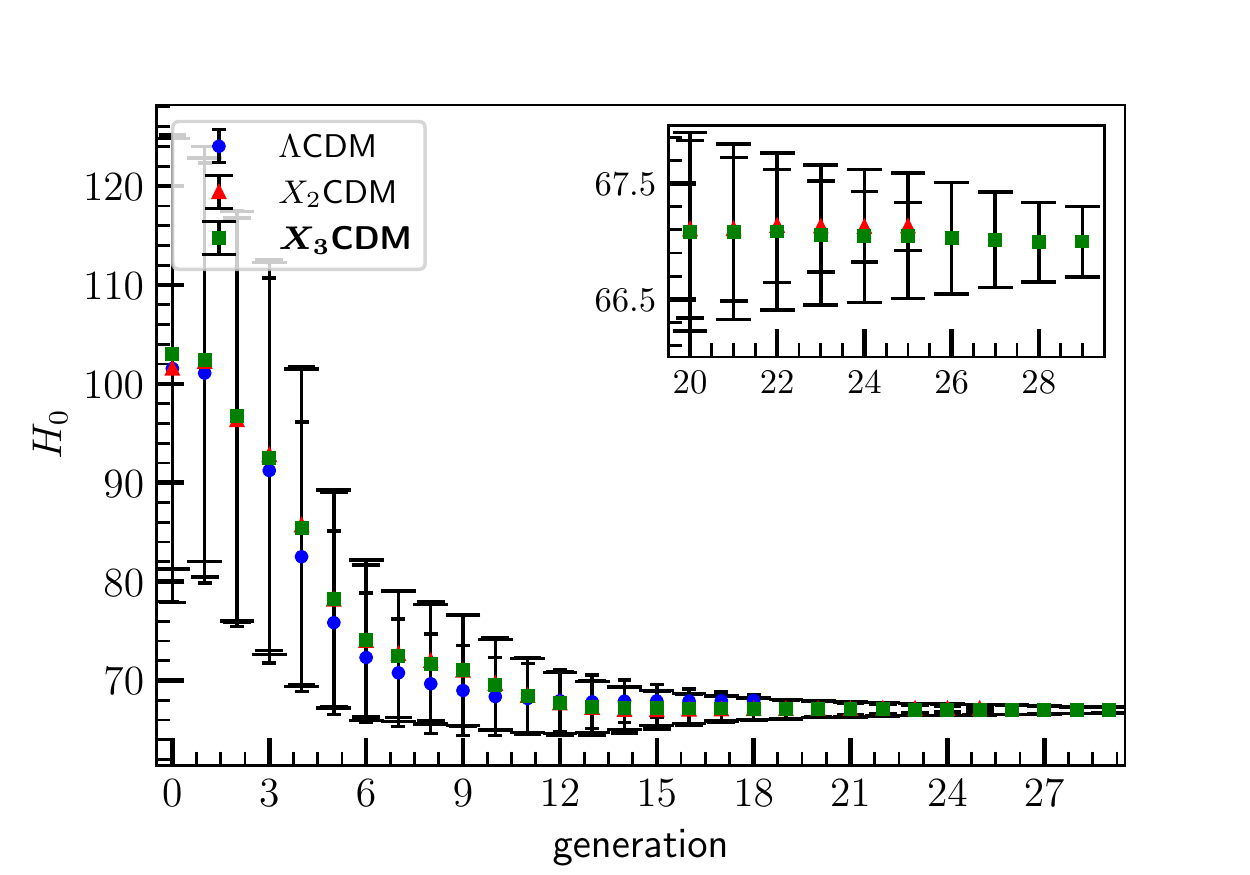}
		}
	\subfigure[ \ matter density ]{
		\includegraphics[width = 0.47 \textwidth]{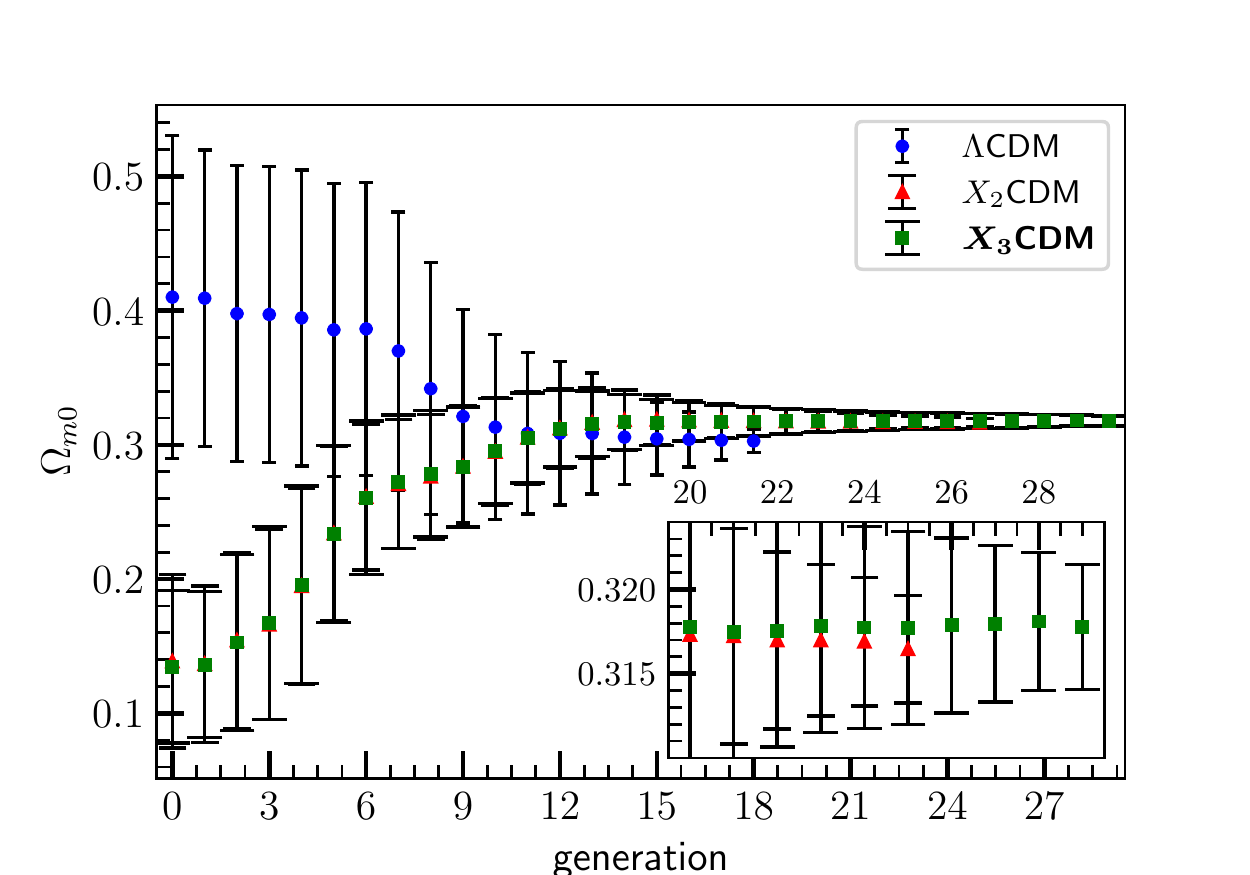}
		}
    \caption{Generational model parameters constrained by the SNe, CC, and BAO data sets. Insets show a zoomed-in of the evolution starting from generation twenty.}
    \label{fig:ABCX23H0om0generational}
\end{figure}

We take a look at the generational evolution of the cosmological parameters in each model to get more insight into the selection. This is shown in Figure \ref{fig:ABCX23H0om0generational}.

We find that the Hubble constants in each model improves with each passing generation, which is not anymore a surprise given ABC. In the beginning, all three models only had quite large uncertainties, in both the Hubble constant and the matter density, which narrowed down continuously as the evolution goes. It is interesting to also pay attention to the matter density evolution in the DDE models, which started out uncharacteristically low. We understand this is due to the DDE models relying on the Planck constraint on the combination $\Omega_{m0} h^2$ to make more meaningful inferences on DE evolution \cite{Aghanim:2018eyx, Grandon:2021nls, Bernardo:2021cxi}. As the Hubble constants started far up in the early evolution, the matter density thus went down in order to keep with a roughly constant $\Omega_{m0} h^2$ prior. Regardless, the cosmological parameters would become more consistent eventually in all models, including $\Lambda$CDM, as shown after about $10$ generations. The parameters of $\Lambda$CDM would then stop from getting measured at generation $19$ when its population shrinks inevitably, while this also eventually happens to $X_2$CDM when it concedes the DDE competition at generation $26$. It is interesting that the DDE model $X_2$CDM started to lose grip of the natural evolution when its matter density fell down, as shown in the inset of Figure \ref{fig:ABCX23H0om0generational}. On the other hand, the cosmological parameters of the naturally selected $X_3$CDM model were as consistent throughout the latter stages of the evolution, that it eventually overcome its remaining competitor.

Table \ref{tab:ABCX23bestfits} shows the cosmological parameters in all three models as determined by the ABC algorithm. Except for the $X_3$CDM model, which won natural evolution, the cosmological parameters were determined by the last surviving population before extinction. 

\begin{table}[h!]
    \centering
    \caption{ABC and MCMC parameter statistics of each model constrained by the SNe, CC, and BAO data sets. For ABC, the $\Lambda$CDM and $X_2$CDM parameters are determined by the last surviving populations.}
    \begin{tabular}{|c|c|c|c|c|c|c|} \hline
    \phantom{gg} algo \phantom{gg} & \phantom{gg} model \phantom{gg} & $H_0$ [km s$^{-1}$Mpc$^{-1}$] & \phantom{gggg} $\Omega_{m0}$ \phantom{gggg} & \phantom{gggg} $x_1$ \phantom{gggg} & \phantom{gggg} $x_2$ \phantom{gggg} & \phantom{gggg} $x_3$ \phantom{gggg} \\ \hline \hline
    \multirow{3}{*}{ABC} & $\Lambda$CDM & $67.0 \pm 0.5$ & $0.30 \pm 0.01$ & $-$ & $-$ & $-$ \\ 
    & $X_2$CDM & $67.1 \pm 0.2$ & $0.316 \pm 0.00{3}$ & $0.80 \pm 0.05$ & $-0.5 \pm 0.2$ & $-$ \\ 
    & \textbf{$\mathbf{X_3}$CDM} & $\mathbf{66.97 \pm 0.02}$ & $\mathbf{0.317 \pm 0.002}$ & $\mathbf{1.11 \pm 0.01}$ & $\mathbf{0.83 \pm 0.04}$ & $\mathbf{-0.9 \pm 0.1}$ \\ \hline \hline
    \multirow{3}{*}{MCMC} & $\Lambda$CDM & $67.9 \pm 0.8$ & $0.304 \pm 0.013$ & $-$ & $-$ & $-$ \\ 
    & $X_2$CDM & $67.0 \pm 0.8$ & $0.312 \pm  0.021$ & $ 0.9 \pm 0.3$ & $-0.1 \pm 1.1$ & $-$ \\ 
    & $X_3$CDM & $66.9\pm 0.8$ & $0.322 \pm 0.022$ & $1.09 \pm 0.16$ & $0.80 \pm 0.52$ & $-0.7 \pm 1.2$ \\ \hline
    \end{tabular}
    \label{tab:ABCX23bestfits}
\end{table}

Notably, this time with the BAO expansion rate data, the cosmological parameters of $\Lambda$CDM turned out to be consistent with Planck \cite{Aghanim:2018eyx}. On the other hand, the Hubble constants and matter densities of the DDE models appeared consistent with each other, at least by $95\%$ confidence {limits}. Both DDE models also exclude the nondynamical DE limit, $\Lambda$CDM, in their estimates, even though their Hubble constants and matter densities are consistent with the CMB measurements. This is clearly a stronger departure to canonical cosmology than previously without the BAO. Now, in all the DE parameters $x_i$ in $X_2$CDM and $X_3$CDM excludes $\Lambda$CDM in their constraints. We also mention that as a consequence of the BAO expansion rate bring{ing in} tighter measurements, we find the cosmological constraints to be {generally} narrower than without the BAO.

Understandably the BAO data relies on $\Lambda$CDM, as the BAO only directly measures $H(z) r_d$, where $r_d$ is the Hubble drag radius. This largely explains why the $\Lambda$CDM ABC best fits, with the addition of the BAO data, agree better with the Planck $\Lambda$CDM measurements \cite{2011MNRAS.416.3017B, Cuceu:2019for}. But on the other end, it is interesting that with the BAO, the DDE parameters $x_i$ found themselves moving further away from the $\Lambda$CDM limit ($x_i = 1$).

\begin{figure}[h!]
    \centering
	\subfigure[ \ Hubble expansion rate ]{
		\includegraphics[width = 0.47 \textwidth]{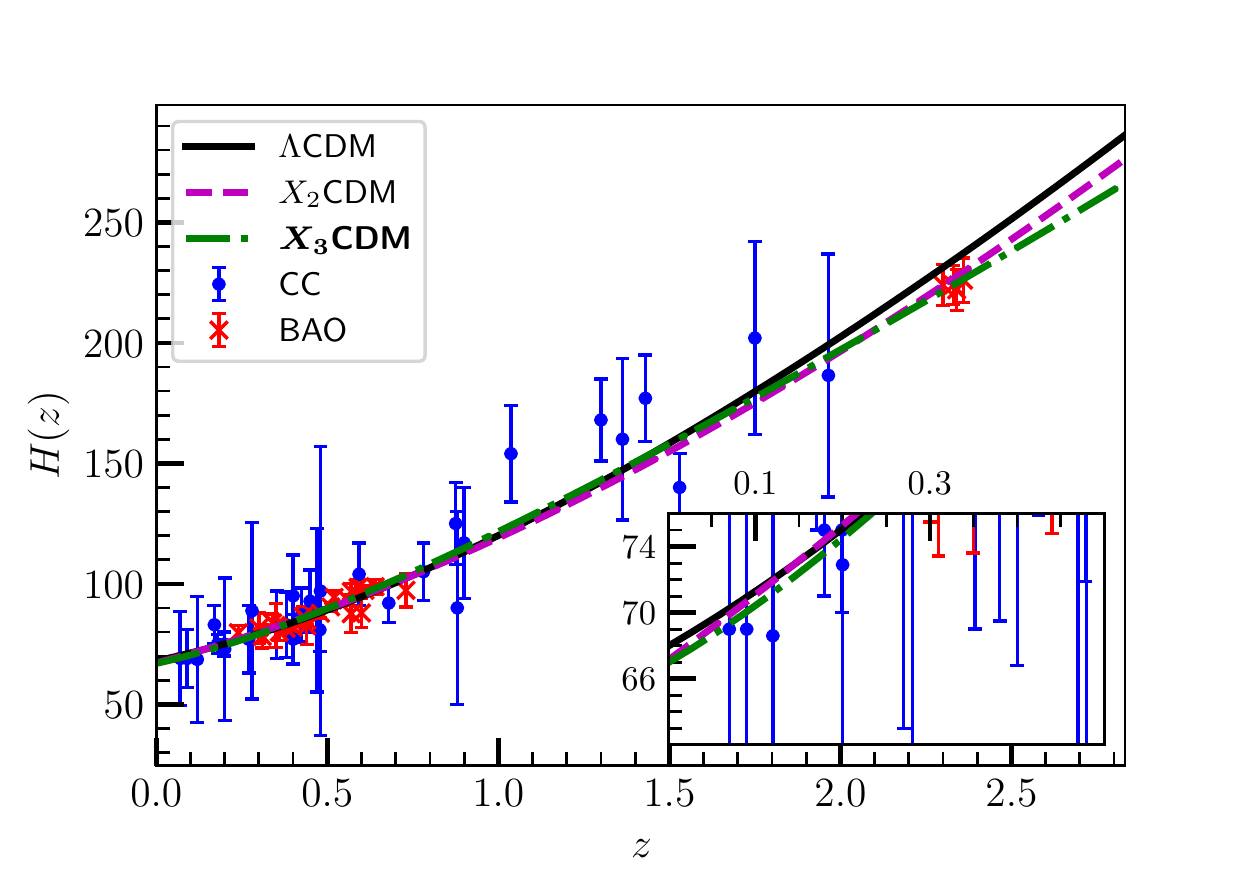}
		}
	\subfigure[ \ SNe brightness ]{
		\includegraphics[width = 0.47 \textwidth]{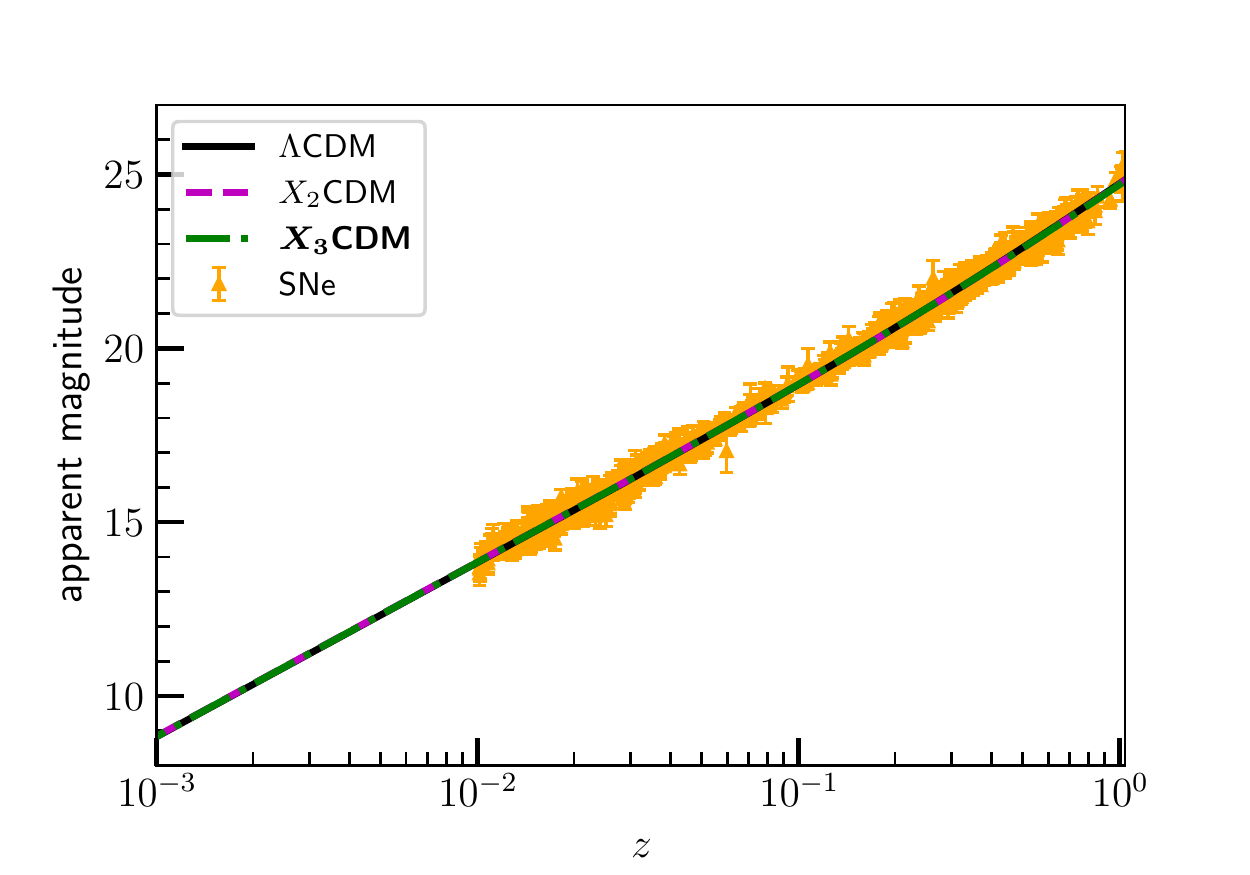}
		}
    \caption{Best fit curves for each model constrained by the SNe, CC, and BAO data sets after thirty generations. For $\Lambda$CDM and $X_2$CDM, the curves are determined by the last surviving population.}
    \label{fig:ABCbestcurves}
\end{figure}

We may draw further insight by looking at the best fit curves in Figure \ref{fig:ABCbestcurves}, resulting from the ABC run with the SNe, CC, and BAO compiled data sets.

We find here now that the BAO is included that both DDE models were preferred by the algorithm as they match with the high redshift Lyman $\alpha$ BAO data at $z \sim 2.3$, in contrast with $\Lambda$CDM which is less flexible and sorely misses these data points. This shows the distance of each member of the $\Lambda$CDM population to be larger, hence the model died out quicker than the DDE ones. The more stringent data set also resulted to the competition between the two DDE models to also be quicker, before $X_3$CDM finally took more control. We see that there is regardless a clear visual distinction between the DDE curves, particularly at the high redshifts outside the scope of the BAO, showcasing how ABC is able to find minute details to distinguish between the competing models.

\subsection{SNe \texorpdfstring{$+$}{} CC \texorpdfstring{$+$}{} BAO \texorpdfstring{$+$}{} \texorpdfstring{$H_0^{\rm P18}$}{}}
\label{subsec:sncbh0}

We move on the ABC results where in addition to considering the late time data set SNe, CC, and BAO, we also take into account a Hubble constant prior. Timely choices for the Hubble constant are the ones reflecting the present Hubble tension, one of cosmology's biggest challenge lately. In this section, we present the results using the Hubble constant prior from Planck \cite{Aghanim:2018eyx}, $H_0^{\rm P18} = 67.4 \pm 0.5$ km s$^{-1}$Mpc$^{-1}$, which makes use of CMB observation, consequently depending on early Universe $\Lambda$CDM physics. 

\begin{figure}[h!]
    \centering
	\subfigure[ \ First sixteen generations ]{
		\includegraphics[width = 0.47 \textwidth]{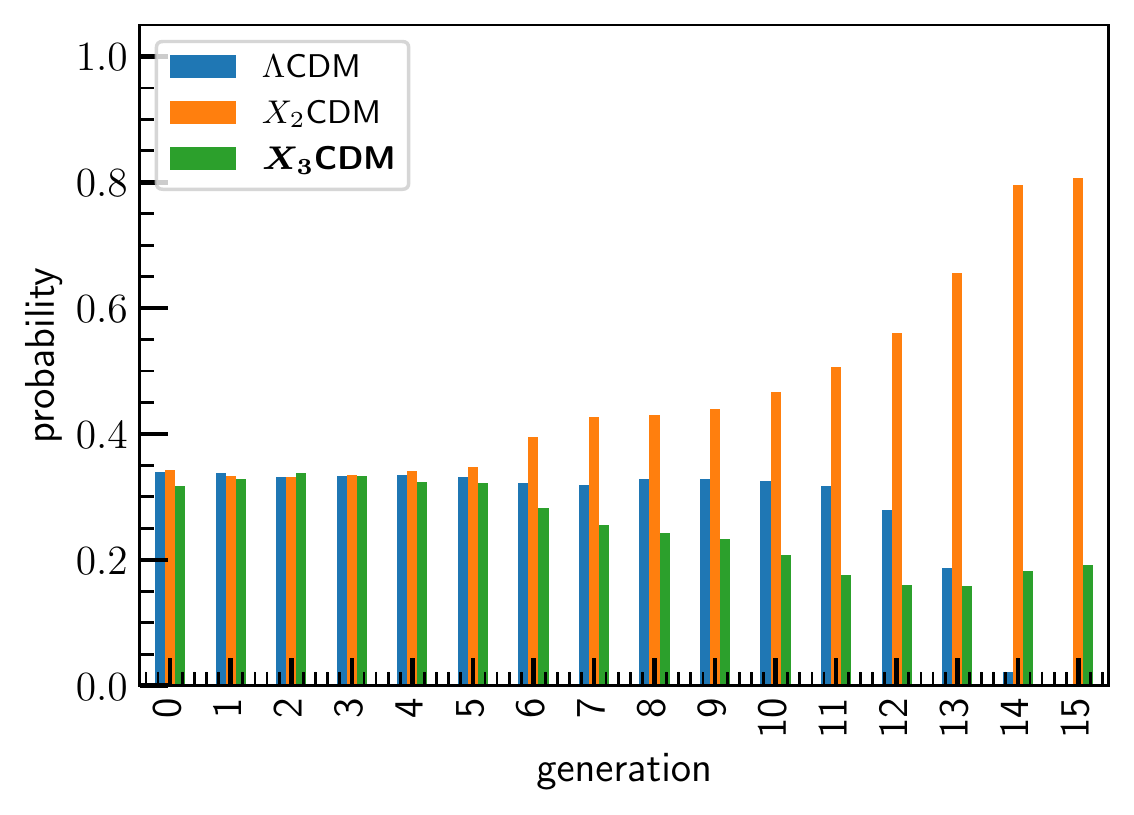}
		}
	\subfigure[ \ After twenty four generations ]{
		\includegraphics[width = 0.47 \textwidth]{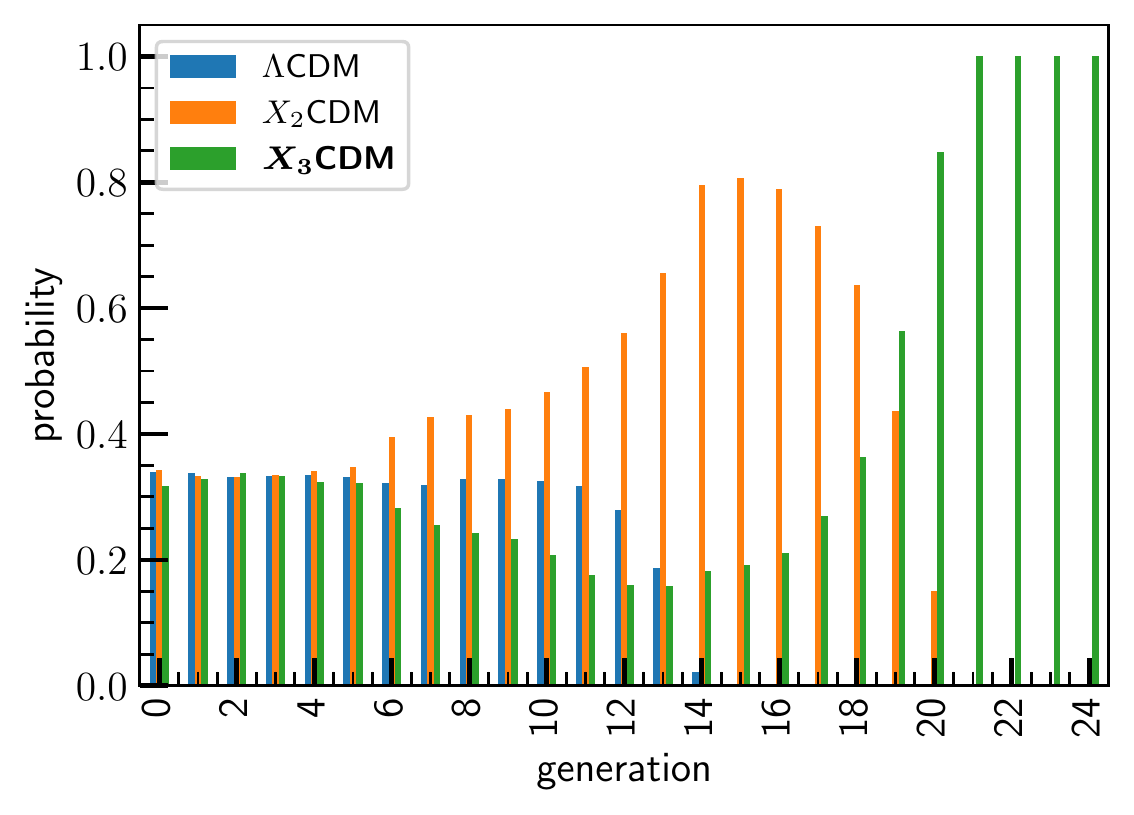}
		}
    \caption{Model posterior evolution constrained by the SNe, CC, and BAO data sets together with the $H_0^\text{P18}$ prior.}
    \label{fig:modelX23post_H0P18}
\end{figure}

The resulting ABC model posterior evolution with the Planck Hubble constant as a prior is shown in Figure \ref{fig:modelX23post_H0P18}.

At this point, it becomes clear that the evolution that ABC conjures up is quite robust, not only whether there is BAO or not, but also with an $H_0$ prior. In the end, it is always the same model that comes out, which in this case happens to be the DDE $X_3$CDM.

Figure \ref{fig:modelX23post_H0P18} shows that the three models started out with as always roughly equal probabilities, as in the previous cases, which will last for about $5$ to $6$ generations. Afterwards, the DDE model $X_2$CDM gains some favor, decreasing the chances for both $\Lambda$CDM and $X_3$CDM in the process. This continues until $\Lambda$CDM goes extinct, or when its population dies out, at generation $15$, leaving the DDE models to compete for their survival. With the two DDE models, it would seem during the early stages that $X_2$CDM may finally come out on top, as the probability for $X_3$CDM declines up to generation $30$. However, at this point, as soon as the DDE models are left by themselves, the $x_3$CDM quickly turns the tables and slowly strengthens its population with each generation. At generation $19$, we see $X_3$CDM overtakes $X_2$CDM, and that at generation $21$, $X_2$CDM goes extinct. The DDE model $X_3$CDM comes out on top again, this time with the Planck Hubble constant as a prior.

\begin{figure}[h!]
    \centering
	\subfigure[ \ Hubble constant ]{
		\includegraphics[width = 0.47 \textwidth]{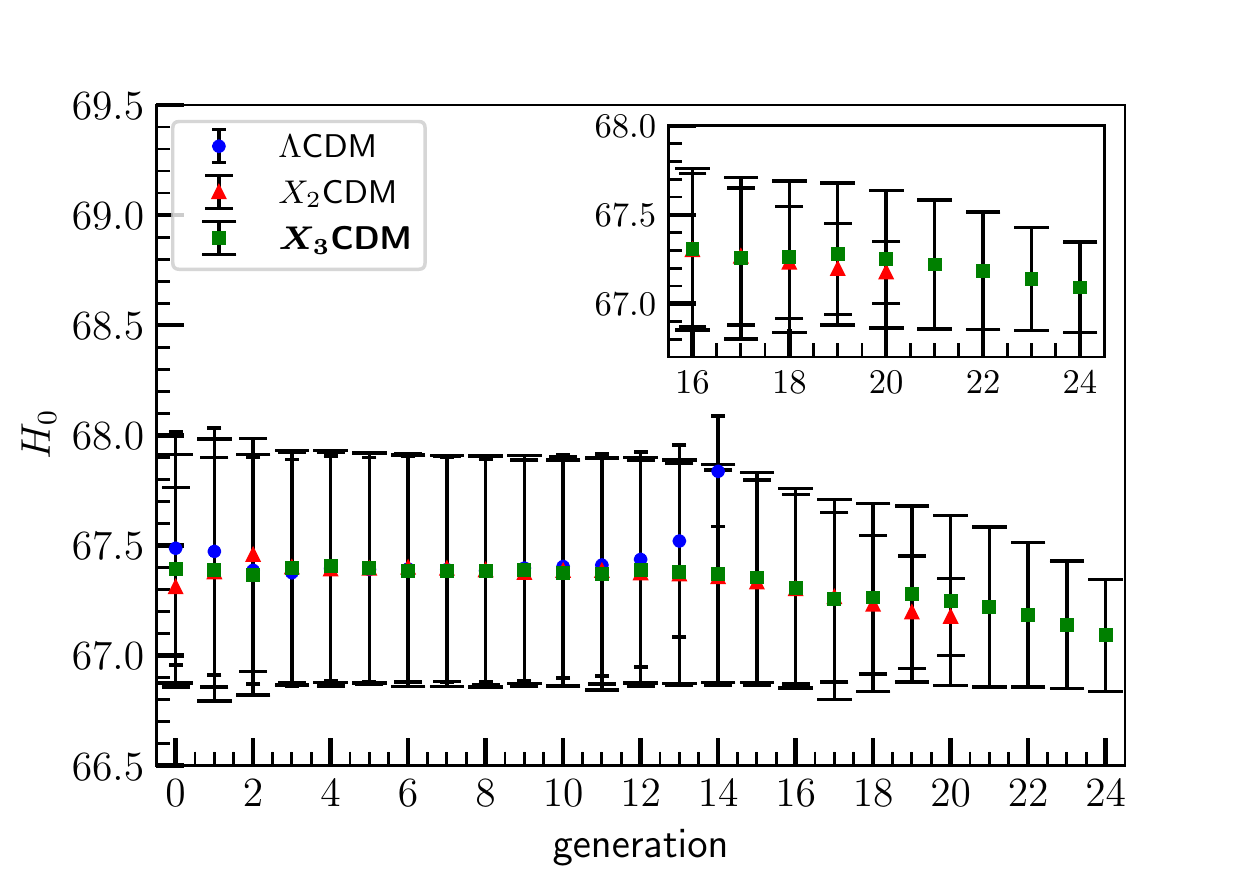}
		}
	\subfigure[ \ matter density ]{
		\includegraphics[width = 0.47 \textwidth]{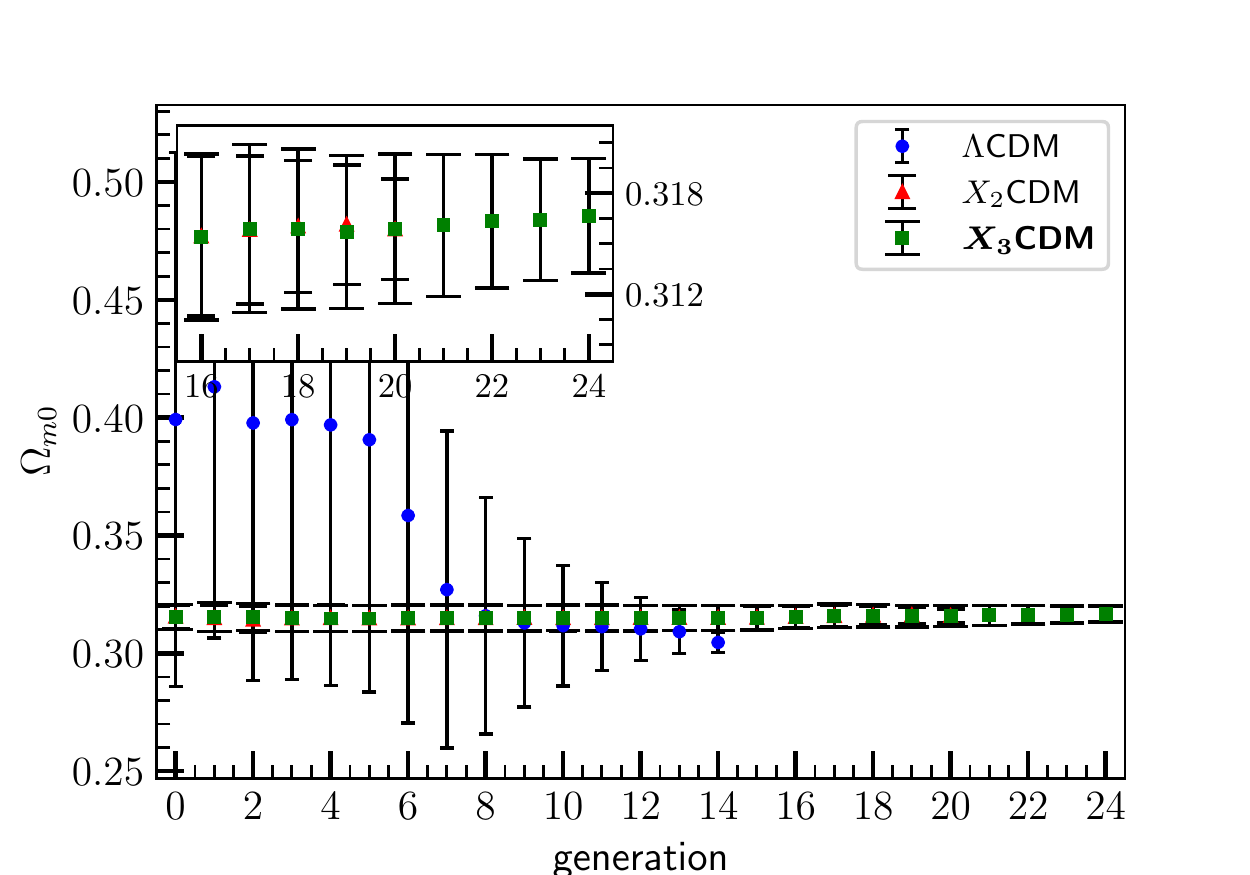}
		}
    \caption{Generational model parameters constrained by the SNe, CC, and BAO data sets together with the $H_0^\text{P18}$ prior. Insets show a zoom in view of the evolution starting from generation sixteen.}
    \label{fig:ABCX23H0om0generational_H0P18}
\end{figure}

It is worth mentioning that the $\Lambda$CDM model went out convincingly quicker with the $H_0^\text{P18}$ than without it. It particularly got extinct by generation $15$ by the ABC evolution selection. Overall the whole evolution to one model went quicker with this more constraining overall data set with a Hubble constant prior. To gain insight into the selection process, we glimpse at the cosmological parameters' evolution shown in Figure \ref{fig:ABCX23H0om0generational_H0P18}.

This makes it clear that now the algorithm takes into account the Hubble constant prior. By about $10$ generations, we see that the Hubble constant within each model matches that of Planck. However, as the selection becomes more competitive, we find the $\Lambda$CDM goes out of contention as soon as its Hubble constant shoots up. This explains why it dropped out quickly in the competition with the Planck Hubble constant prior. The remainder of the competition thus became tighter between the DDE models, as their Hubble constants and matter densities are practically consistent throughout. The reason for this is that both DDE models rely on the Planck constraint $\Omega_{m0} h^2$ which is directly measured by the CMB damping tail. Now that $H_0^\text{P18}$ is preferred, in principle both Hubble constants and matter densities become determined for both models. It is up to the DE parameters $x_i$ to match with the data, and so fight for their population's survival. The inset of Figure \ref{fig:ABCX23H0om0generational_H0P18} reveals that the matter density also played a role in deciding the stronger model by evolution standards. We however shall see this more clearly using a different prior for the Hubble constant.

Table \ref{tab:ABCX23bestfits_H0P18} shows the ABC estimates of the cosmological parameters, with the MCMC counterparts, for each model with the Planck Hubble constant prior.

\begin{table}[h!]
    \centering
    \caption{ABC and MCMC parameter statistics of each model constrained by the SNe, CC, and BAO data sets together with the $H_0^\text{P18}$ prior. For $\Lambda$CDM and $X_2$CDM the parameters are determined by the last surviving populations.}
    \begin{tabular}{|c|c|c|c|c|c|c|} \hline
    \phantom{gg} algo \phantom{gg} & \phantom{gg} model \phantom{gg} & $H_0$ [km s$^{-1}$Mpc$^{-1}$] & \phantom{gggg} $\Omega_{m0}$ \phantom{gggg} & \phantom{gggg} $x_1$ \phantom{gggg} & \phantom{gggg} $x_2$ \phantom{gggg} & \phantom{gggg} $x_3$ \phantom{gggg} \\ \hline \hline
    \multirow{3}{*}{ABC} & $\Lambda$CDM & $67.8 \pm 0.3$ & $0.305 \pm 0.004$ & $-$ & $-$ & $-$ \\
    & $X_2$CDM & $67.2 \pm 0.2$ & $0.316 \pm 0.00{3}$ & $0.80 \pm 0.05$ & $-0.5 \pm 0.2$ & $-$ \\
    & \textbf{$\mathbf{X_3}$CDM} & $\mathbf{66.97 \pm 0.01}$ & $\mathbf{0.317 \pm 0.002}$ & $\mathbf{1.11 \pm 0.01}$ & $\mathbf{0.83 \pm 0.04}$ & $\mathbf{-0.9 \pm 0.1}$ \\ \hline \hline
    \multirow{3}{*}{MCMC} & $\Lambda$CDM & $67.5 \pm 0.4$ & $0.31 \pm  0.01$ & $-$ & $-$ & $-$ \\
    & $X_2$CDM & $67.3\pm 0.4$ & $0.31 \pm 0.02$ & $0.9 \pm 0.3$ & $-0.05 \pm 1.1$ & $-$ \\
    & $X_3$CDM & $67.2 \pm 0.4$ & $0.32 \pm 0.02$ & $1.08 \pm 0.16$ & $0.81 \pm 0.52$ & $-0.6 \pm 1.2$ \\ \hline
    \end{tabular}
    \label{tab:ABCX23bestfits_H0P18}
\end{table}

One observation stands out: with or without the $H_0^\text{P18}$ prior, the ABC determined cosmological parameters of the $X_3$CDM are consistent given the SNe, CC, and BAO compiled data sets. This also seemingly is the case with $X_2$CDM, except that the ABC estimates of $x_2$ disagree when the Planck Hubble constant prior was considered. We remind again that as far as ABC goes, only the cosmological parameters of the naturally evolved $X_3$CDM are meaningful, and that we present the corresponding parameters of $\Lambda$CDM and $X_2$CDM only for discussion's sake. It is also worth mentioning that the $\Lambda$CDM cosmological parameters are consistent with Planck, which maybe trivial given that imposing a Hubble constant prior on this model just leaves only one more parameter to be constrained.

\begin{figure}[h!]
    \centering
	\subfigure[ \ Hubble expansion rate ]{
		\includegraphics[width = 0.47 \textwidth]{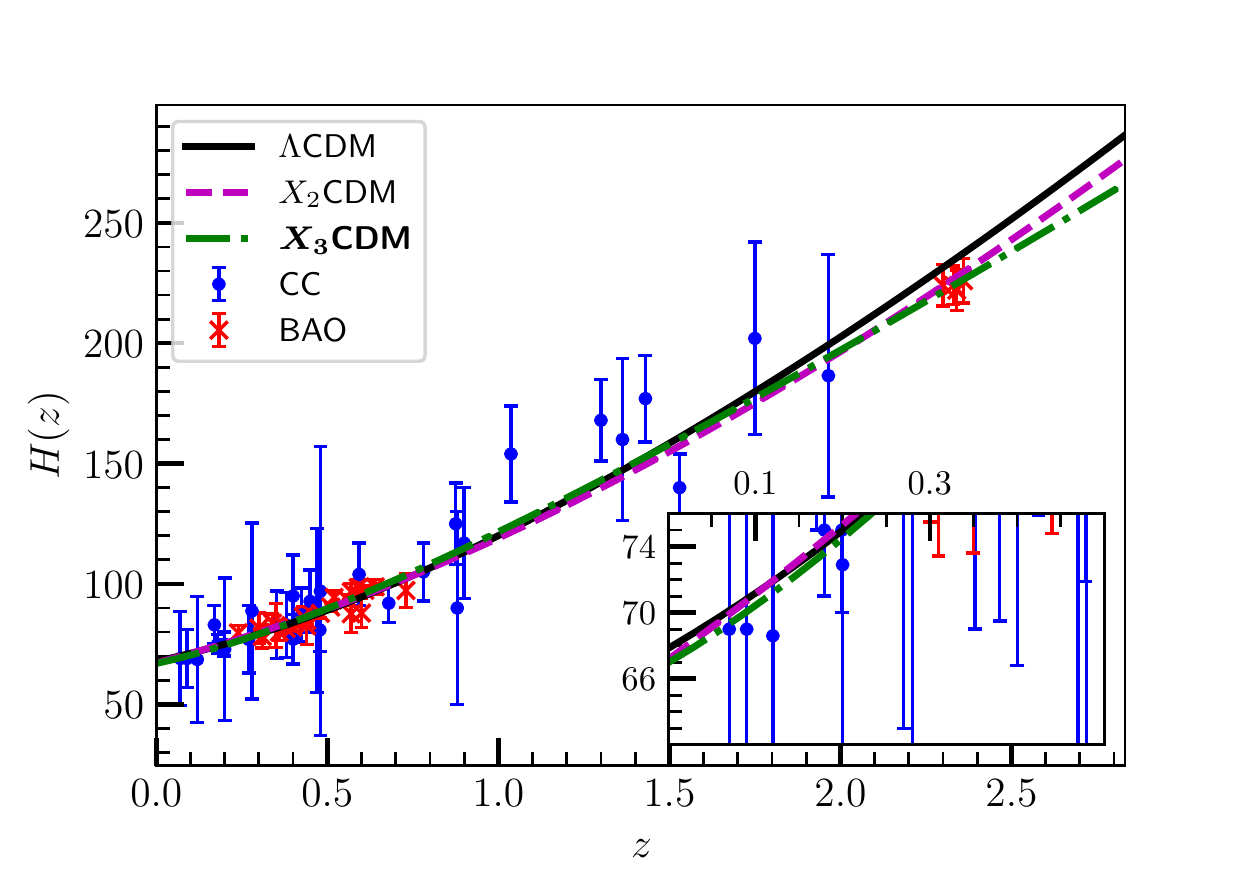}
		}
	\subfigure[ \ SNe brightness ]{
		\includegraphics[width = 0.47 \textwidth]{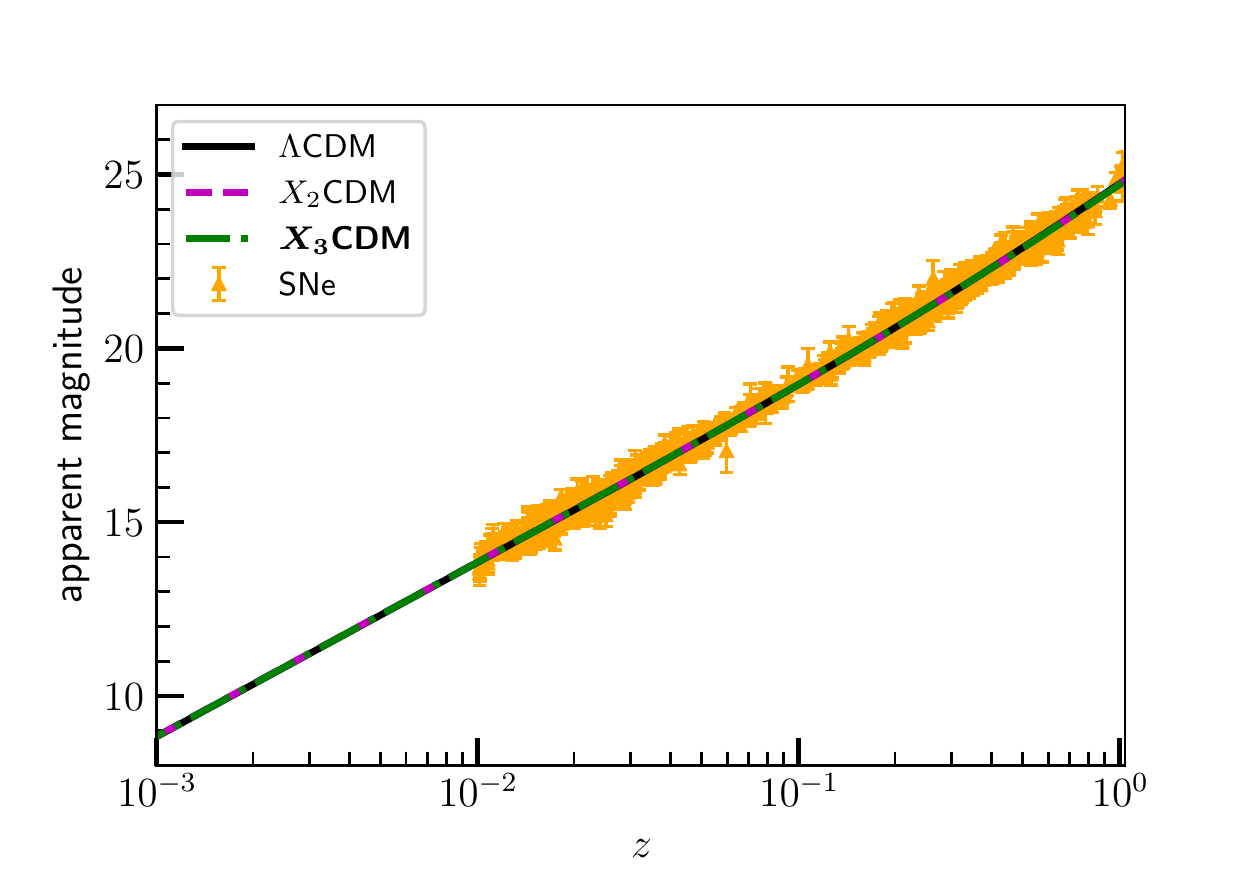}
		}
    \caption{Best fit curves for each model constrained by the SNe, CC, and BAO data sets together with the $H_0^\text{P18}$ prior after forty generations. For $\Lambda$CDM and $X_2$CDM, the curves are determined by the last surviving population.}
    \label{fig:ABCbestcurves_H0P18}
\end{figure}

We again receive insight as to how the competition went through by looking at the best fit curves in each model as determined by ABC. This is shown in Figure \ref{fig:ABCbestcurves_H0P18}.

As it turned out, with or without the $H_0^\text{P18}$ prior, the Hubble constant estimates in each DDE model turned out consistent with each other, as revealed more closely by the inset. That is, a low value of the Hubble constant is preferred, intriguingly consistent with the Planck data. We also look past the best fit SNe magnitude curves as the SNe data set has already done its job in restricting the curves quite tightly, and instead focus on the Hubble expansion rate curves where there is more noise in the data. As in the previous case without the $H_0$ prior, we find that it is the BAO Lyman $\alpha$ data points that segregate the competition between $\Lambda$CDM and the two DDE models. The tough competition between the DDE models showcase ABC's potential in finding minute distinctions between the observation and models. We can regardless distinguish between the two DDE models by looking at large redshifts, $z \gtrsim 2.3$, beyond the Lyman alpha BAO data points. We find in this region that the surviving $X_3$CDM model predicts a slightly lower expansion rate compared to that of $X_2$CDM and $\Lambda$CDM.

\subsection{SNe \texorpdfstring{$+$}{} CC \texorpdfstring{$+$}{} BAO \texorpdfstring{$+$}{} \texorpdfstring{$H_0^{\rm R22}$}{}}
\label{subsec:sncbh0}

Finally we consider the ABC estimation of the cosmological parameters with the Hubble constant from supernovae \cite{Riess:2021jrx}, $H_0^{\rm R22} = 73.30 \pm 1.04$ km s$^{-1}$Mpc$^{-1}$, taken as a prior in the analysis. This is relatively high compared with the CMB estimate, reflecting the Hubble tension, and will turn out to be quite tricky with the ABC as we show.

\begin{figure}[h!]
    \centering
	\subfigure[ \ First thirty generations ]{
		\includegraphics[width = 0.47 \textwidth]{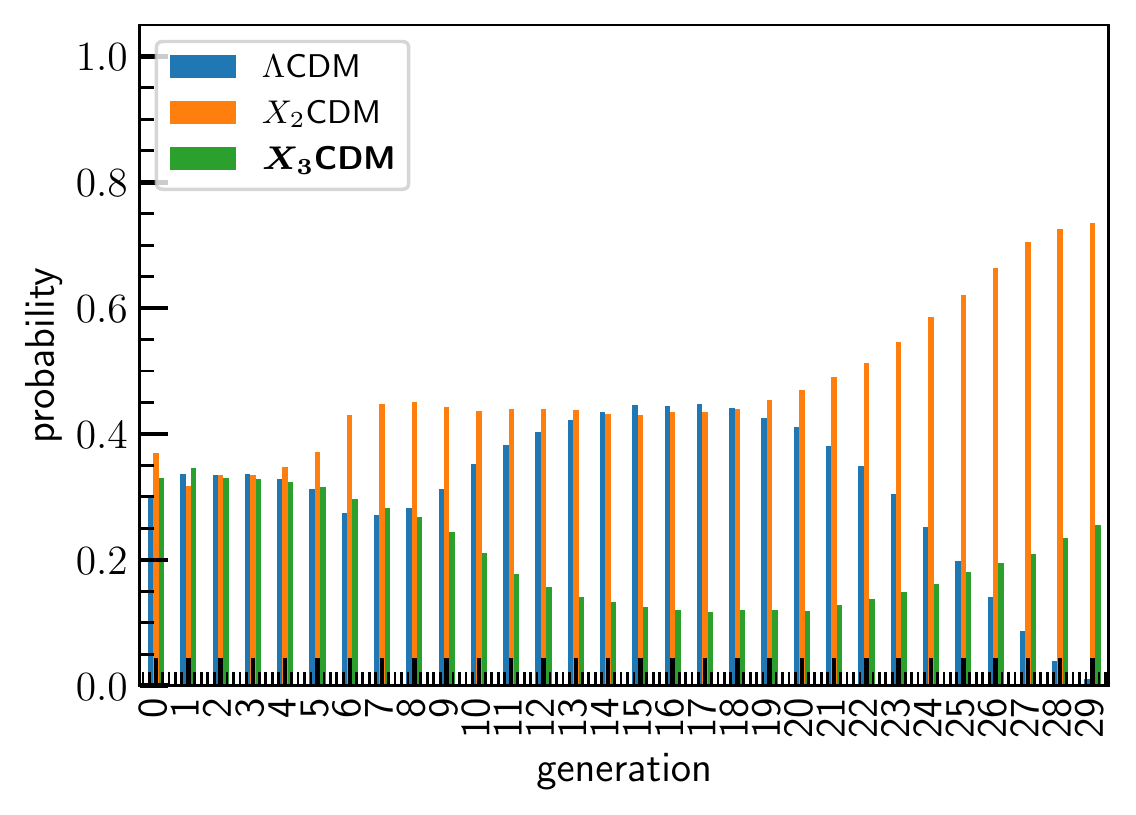}
		}
	\subfigure[ \ After forty five generations ]{
		\includegraphics[width = 0.47 \textwidth]{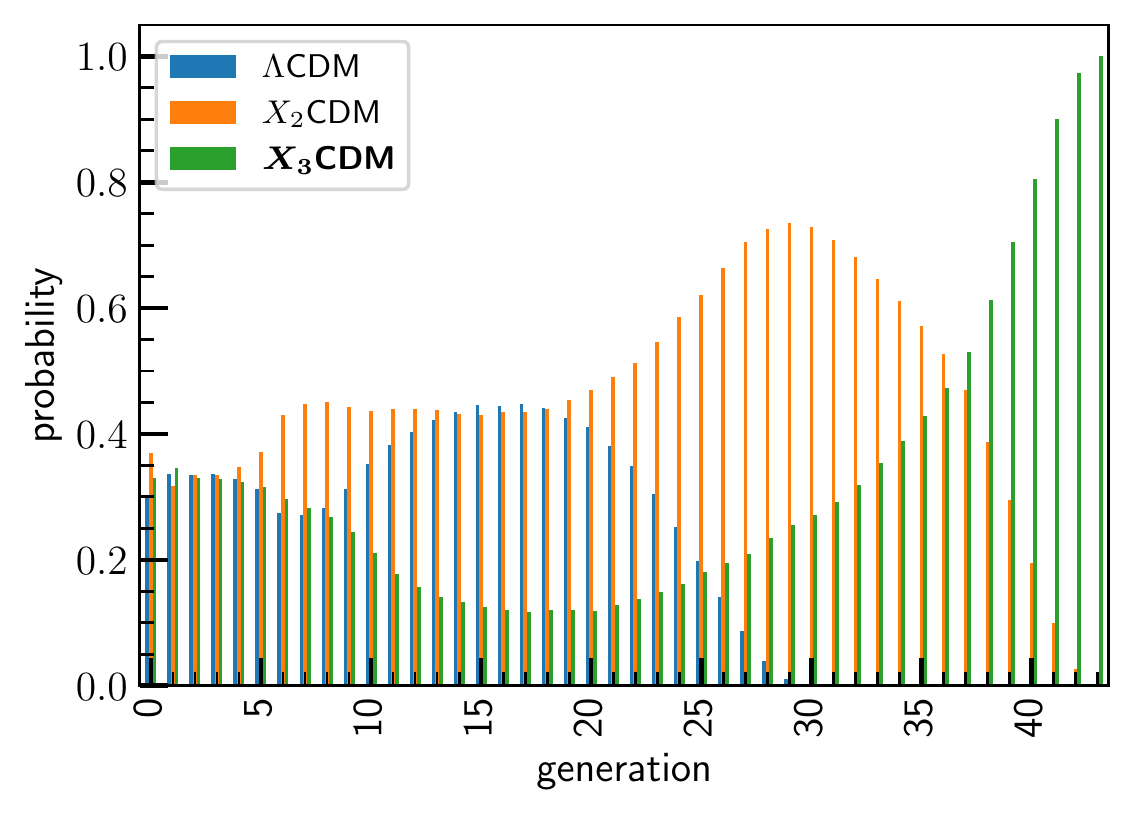}
		}
    \caption{Model posterior evolution constrained by the SNe, CC, and BAO data sets together with the $H_0^\text{R22}$ prior.}
    \label{fig:modelX23post_H0R22}
\end{figure}

The model posterior evolution is shown in Figure \ref{fig:modelX23post_H0R22}. 

\begin{figure}[h!]
    \centering
	\subfigure[ \ Hubble constant ]{
		\includegraphics[width = 0.47 \textwidth]{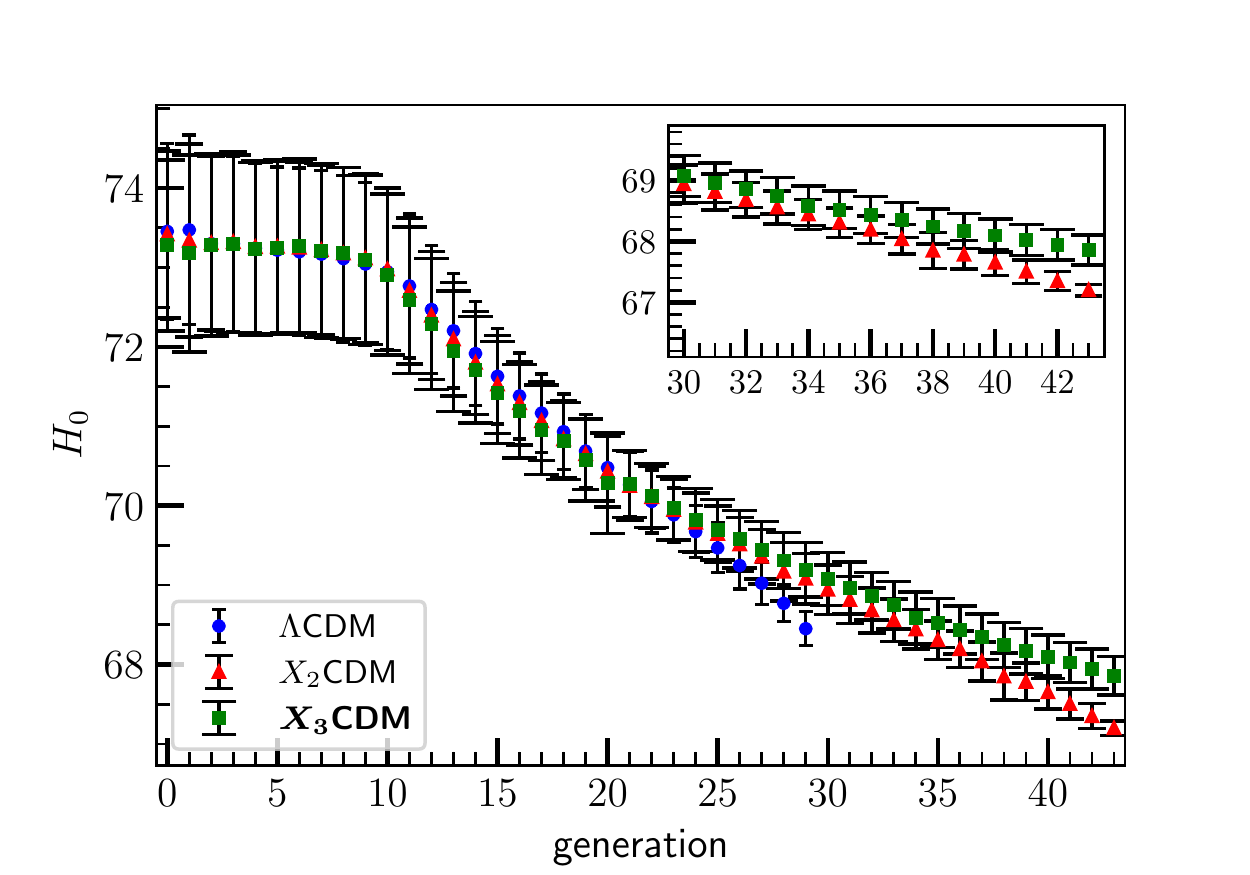}
		}
	\subfigure[ \ matter density ]{
		\includegraphics[width = 0.47 \textwidth]{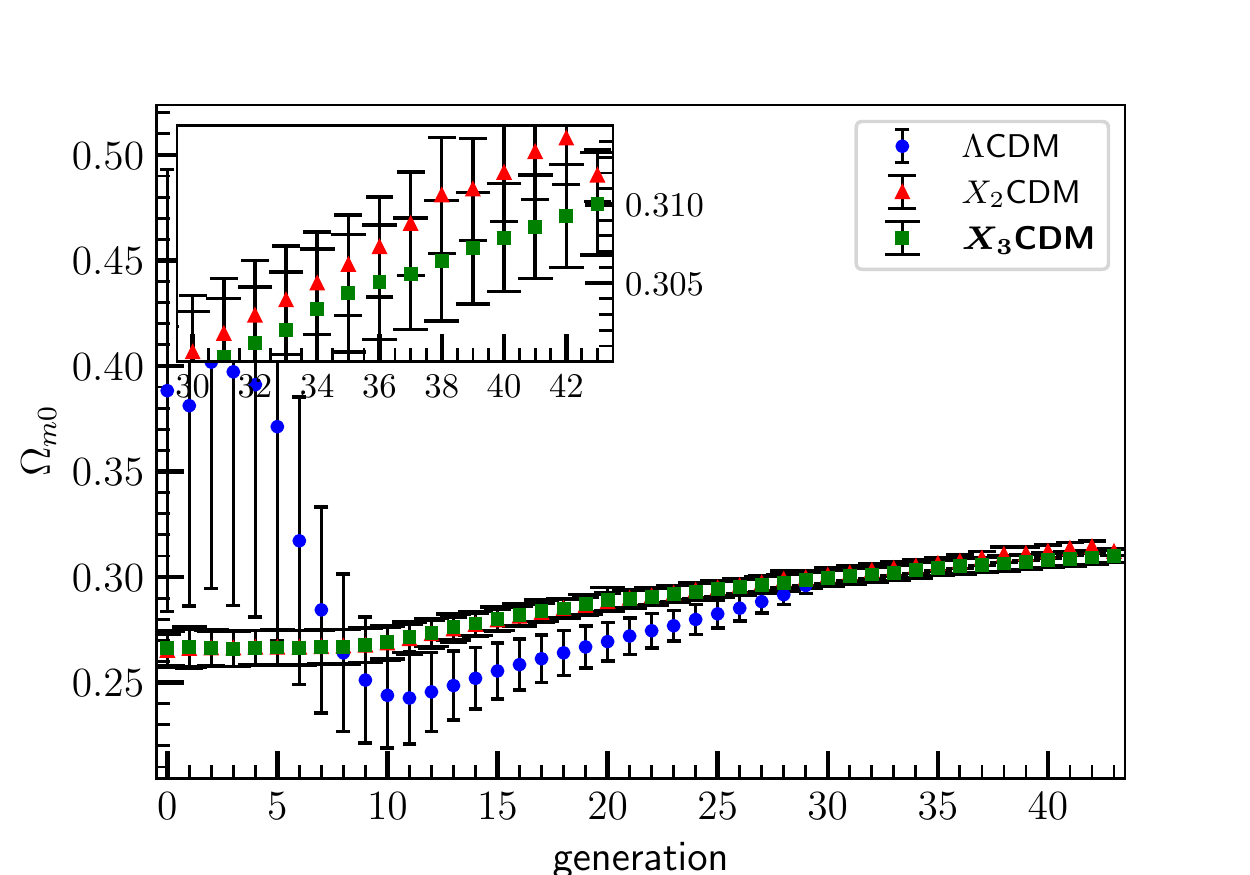}
		}
    \caption{Generational model parameters constrained by the SNe, CC, and BAO data sets together with the $H_0^\text{R22}$ prior. Insets show a zoom in view of the evolution starting from generation thirty.}
    \label{fig:ABCX23H0om0generational_H0R22}
\end{figure}

We see here that $X_2$CDM dominated the early evolution as it did in the previous cases. Then, after $29$ generations, the $\Lambda$CDM model collapsed, leaving once more the DDE models to compete for survival. Note that it took longer this time for the $\Lambda$CDM to gone out of the evolution. What happened after was also quite intriguing as the battle for natural selection dragged on for a longer period until generation $42$. Then, the $X_2$CDM model turned unfit and its population eventually collapsed, leaving behind the robust stronger model, $X_3$CDM. We can understand the evolution better by looking at the model parameters' evolution in Figure \ref{fig:ABCX23H0om0generational_H0R22}.

This is showing that the best estimate Hubble constant, $H_0 \lesssim 68$ km s$^{-1}$ Mpc$^{-1}$, in each model have gone quite too far below the prior $H_0 \sim 73$ km s$^{-1}$ Mpc$^{-1}$ during the course of the evolution. We takeaway from this that the ABC together with the late time data sets we consider here prefer low values of the Hubble constant that are consistent with Planck. This shrinks down the population of each model while the evolution drags on as the distance of each model to the observation and the priors becomes overall larger since $H_0$ drops. On the other end, the matter density evolved quite as expected at first, with the one by $\Lambda$CDM nearly adjusting to a value consistent with the other cases. However, the $\Lambda$CDM model drops out before its matter density can keep up with the DDEs. We also mention that for the DDEs, since the combination $\Omega_{m0} h^2$ is constrained by the CMB damping tail, specifying $H0$ practically constrains the matter density $\Omega_{m0}$. This explains the tight estimate and slow increase in the matter density estimates per generation in Figure \ref{fig:ABCX23H0om0generational_H0R22}, that is, while $H_0$ drops to a preferably low value consistent with the data, $\Omega_{m0}$ adjusts upward in the parameter space.

As in the previous cases, it is of course possible to take the best estimates from the models' last populations. In this case with the $H_0^\text{R22}$ prior, we put in the ABC statistics per model as well as their MCMC counterparts in Table \ref{tab:ABCX23bestfits_H0R22}.

\begin{table}[h!]
    \centering
    \caption{ABC and MCMC parameter statistics of each model constrained by the SNe, CC, and BAO data sets together with the $H_0^\text{R22}$ prior. For $\Lambda$CDM and $X_2$CDM the parameters are determined by the last surviving populations.}
    \begin{tabular}{|c|c|c|c|c|c|c|} \hline
    \phantom{gg} algo \phantom{gg} & \phantom{gg} model \phantom{gg} & $H_0$ [km s$^{-1}$Mpc$^{-1}$] & \phantom{gggg} $\Omega_{m0}$ \phantom{gggg} & \phantom{gggg} $x_1$ \phantom{gggg} & \phantom{gggg} $x_2$ \phantom{gggg} & \phantom{gggg} $x_3$ \phantom{gggg} \\ \hline \hline
    \multirow{3}{*}{ABC} & $\Lambda$CDM & $68.5 \pm 0.2$ & $0.296 \pm 0.004$ & $-$ & $-$ & $-$ \\
    & $X_2$CDM & $67.2 \pm 0.1$ & $0.312 \pm 0.00{2}$ & $0.85 \pm 0.02$ & $-0.26 \pm 0.07$ & $-$ \\
    & \textbf{$\mathbf{X_3}$CDM} & $\mathbf{67.9 \pm 0.2}$ & $\mathbf{0.3100 \pm 0.00{3}}$ & $\mathbf{1.05 \pm 0.04}$ & $\mathbf{0.8 \pm 0.1}$ & $\mathbf{-0.7 \pm 0.2}$ \\ \hline \hline
    \multirow{3}{*}{MCMC} & $\Lambda$CDM & $67.0 \pm 0.6$ & $0.28\pm 0.01$ & $-$ & $-$ & $-$ \\
    & $X_2$CDM & $69.6 \pm 0.7$ & $0.29 \pm 0.02$ & $0.78 \pm 0.28$ & $-0.2 \pm 1.0$ & $-$ \\
    & $X_3$CDM & $69.6 \pm 0.7$ & $ 0.30 \pm 0.02$ & $0.97\pm 0.15$ & $0.6 \pm 0.5$ & $-0.6 \pm 1.1$ \\ \hline
    \end{tabular}
    \label{tab:ABCX23bestfits_H0R22}
\end{table}

This echoes that the $H_0$ estimates from the ABC wander in the low $H_0$ region more agreeable with the Planck estimate. The values presented in Table \ref{tab:ABCX23bestfits_H0R22} as a matter of fact can be seen to be fairly consistent with the ones obtained with instead the Planck prior for $H_0$. Since the Hubble constant estimates are quite far from the prior $H_0^\text{R22}$, this explains why the ABC eventually shrinks $\Lambda$CDM and $X_2$CDM, leaving the $X_3$CDM naturally selected. Eventually the overall distance becomes too large for the less flexible models that their populations die out. Nonetheless, one thing we can straighten out is that ABC consistently prefers the low values of $H_0$, more consistent with the CMB, than high ones from distance ladder anchors. We find this to be true independent of the model, by ABC standards.

\begin{figure}[h!]
    \centering
	\subfigure[ \ Hubble expansion rate ]{
		\includegraphics[width = 0.47 \textwidth]{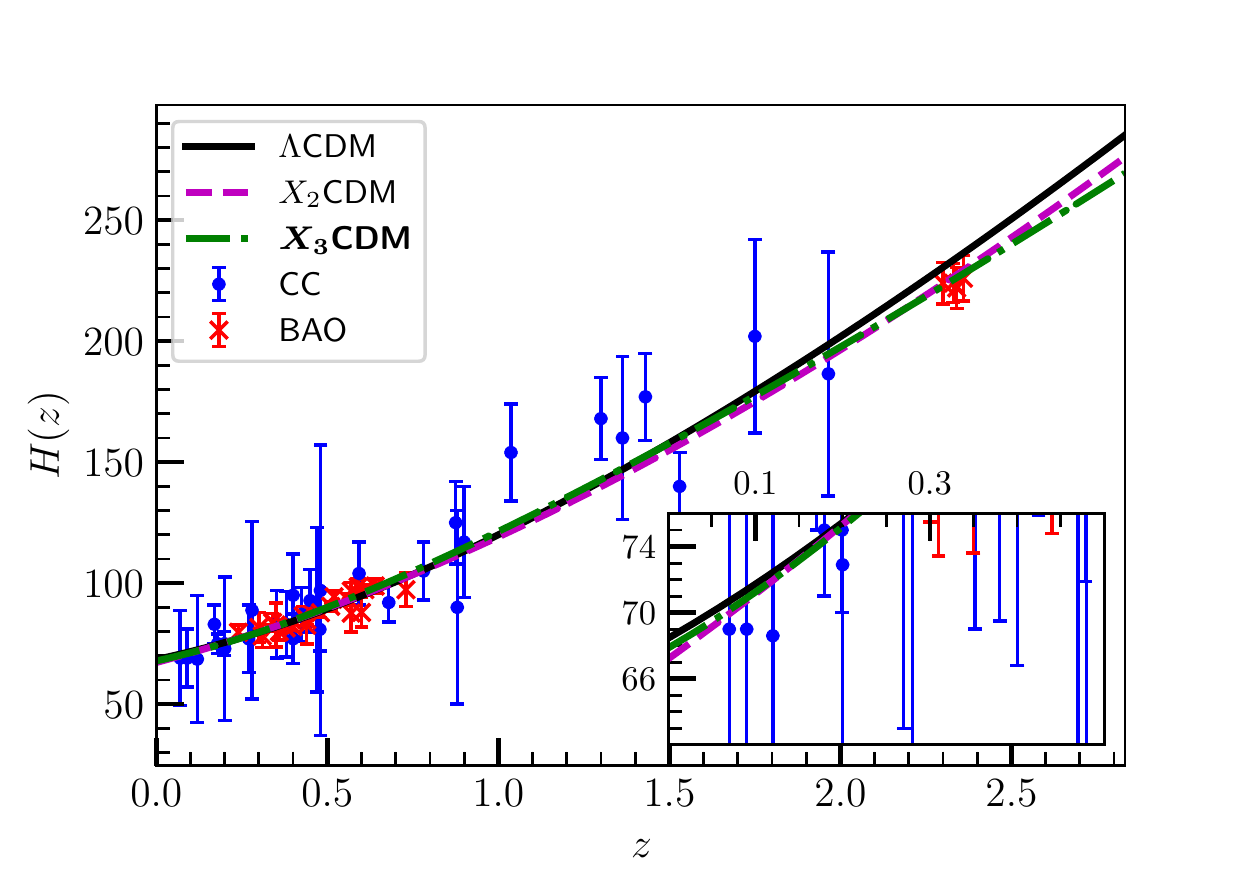}
		}
	\subfigure[ \ SNe brightness ]{
		\includegraphics[width = 0.47 \textwidth]{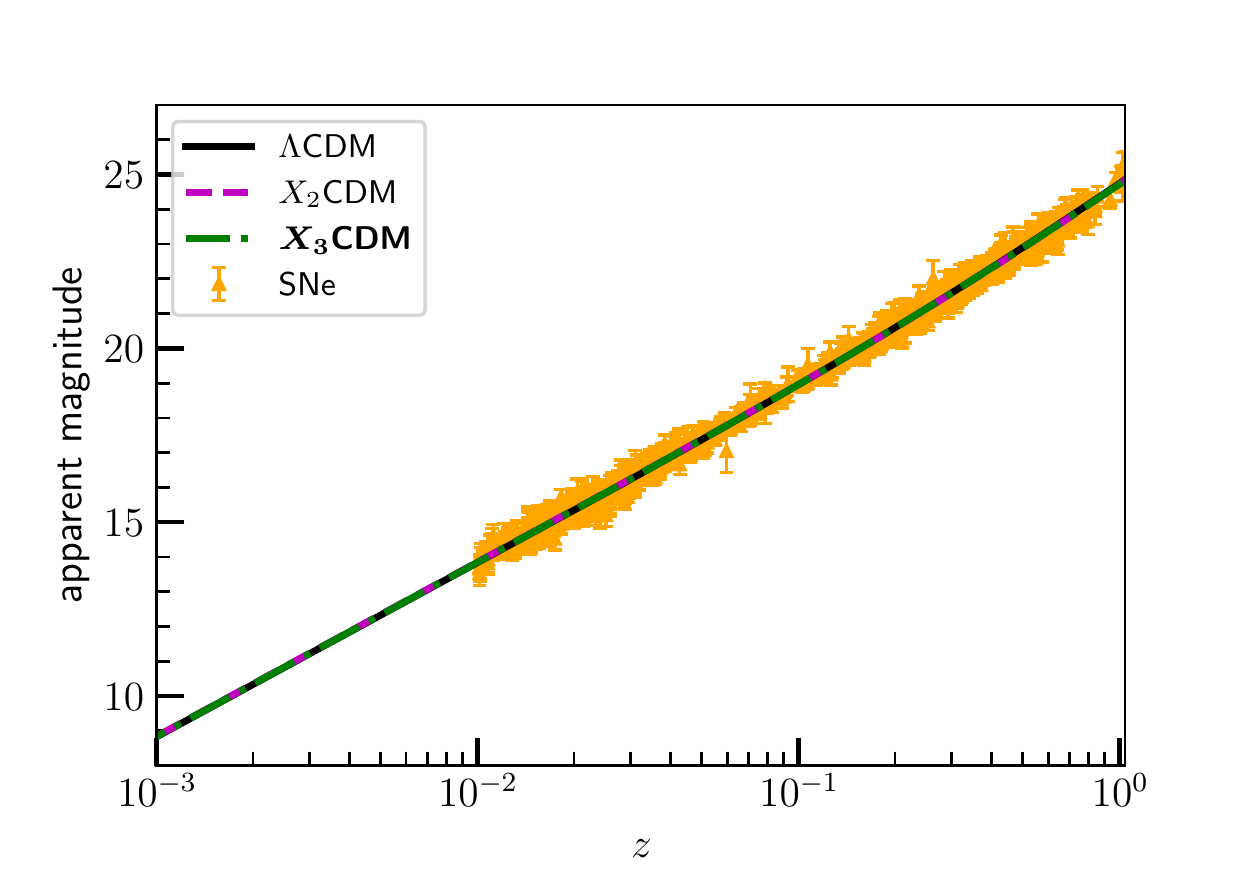}
		}
    \caption{Best fit curves for each model constrained by the SNe, CC, and BAO data sets together with the $H_0^\text{R22}$ prior after forty five generations. For $\Lambda$CDM and $X_2$CDM, the curves are determined by the last surviving population.}
    \label{fig:ABCbestcurves_H0R22}
\end{figure}

The best fit curves, obtained from each model are shown in Figure \ref{fig:ABCbestcurves_H0R22}.

Again, we see how robust the ABC results are. Overlooking the SNe brightness best fits which are distinguishable, we find that the differentiator between the nondynamical $\Lambda$CDM and the DDE were clearly the Lyman $\alpha$ BAO points. Also consistent with all the previous results, we find that the naturally selected model, $X_3$CDM, has a lower expansion rate at the higher redshifts, $z \gtrsim 2.3$, compared with $X_2$CDM and $\Lambda$CDM. We obtained this regardless of the various splices of the data we use and of the Hubble constant priors.

\section{Discussion}
\label{sec:discussion}

Looking at the results (Tables \ref{tab:ABCX23bestfits_noBAO}, \ref{tab:ABCX23bestfits}, \ref{tab:ABCX23bestfits_H0P18}, and \ref{tab:ABCX23bestfits_H0R22}), we notice that the best fit to the data persistently prefers a value greater than one for the parameter $x_1$ for the cubic parametrization. This means that the reconstructed function $X(z)$ grows slightly at low $z$ and then starts to decrease until it takes even negative values (see the sign of $x_3$ in the tables above). This behavior was first noticed in \cite{Cardenas:2014jya} and \cite{Magana:2014voa}, and at that time was related to the `cosmic slowing down of acceleration' phenomena studied in \cite{Shafieloo:2009ti}.

We have viewed dark energy from a biology inspired perspective by means of the ABC algorithm and several late time cosmological observations. Our results were robust in speaking out dynamical dark energy as preferred over to its nondynamical counterpart, the standard $\Lambda$CDM cosmological model. We found this holds regardless if we take into consideration the BAO data points and with or without Hubble constant priors. Understandably, the result is controversial as it stands on the other side of standard cosmology, which is supported by numerous astrophysical observations beyond just the late time Universe. So we must look at this with skeptical eyes. But all the same we acknowledge that noncanonical methods such as ABC add new flavor in cosmological analysis or at least give a different path to the same results, consequently strengthening the prevailing theory. We for one found that regardless of the choice of the models, data sets and priors we considered, the Hubble constant is always consistent with the Planck data.

As it stands, our ABC results are supported by the traditional MCMC, thereby adding to the case of DDE being a more statistically favored model of the late Universe. In contrast, ABC views the selection through a model space that assigns a statistical identity to $\Lambda$CDM together with its DDE parametric extensions. This is made possible by the ABC which sidesteps the reliance on a likelihood function and approximates the posterior in an increasingly precise series of generations.

\begin{figure}[h!]
    \centering
    \includegraphics[width = 0.9 \textwidth]{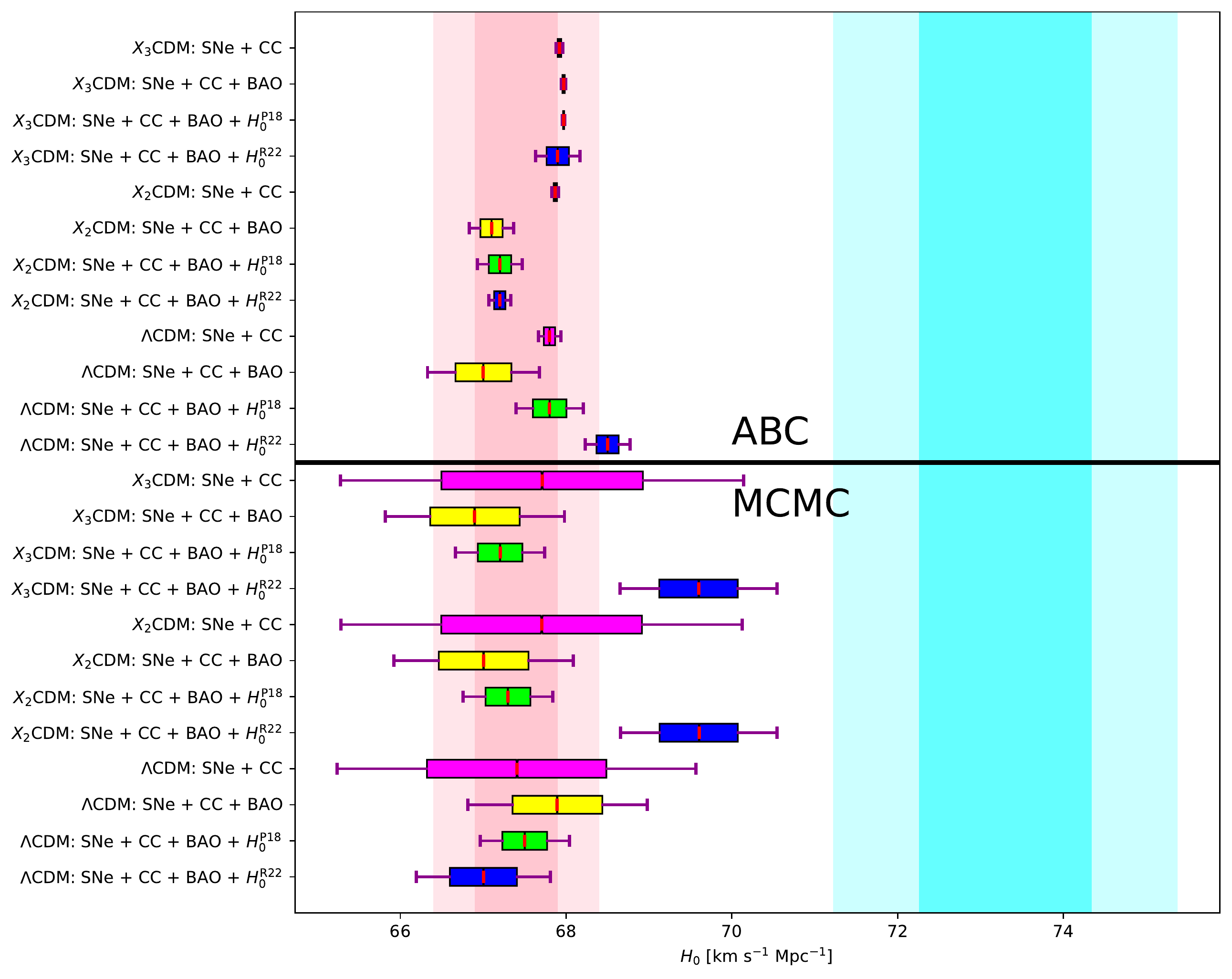}
    \caption{Whisker plot of the Hubble constants ($95\%$ confidence) obtained in this work with reference to the Planck \cite{Aghanim:2018eyx} (pink band) and SH0ES \cite{Riess:2021jrx} (blue band) estimates. More accurately, the box represents the first quartile to the third quartile of the $H_0$ distribution from the median, while the whiskers extend from the box by $0.5$ times the interquartile range.}
    \label{fig:whiskerH0}
\end{figure}

The agreement between MCMC and ABC parameter estimates is drawn out in Figure \ref{fig:whiskerH0} where the various Hubble constants obtained in the work are presented together with reference values from the Planck \cite{Aghanim:2018eyx} and SH$0$ES collaborations \cite{Riess:2021jrx}. 

This reveals the impact of ABC on the Hubble constant, and indirectly on the other cosmological parameters. When comparing the two methods, we always find that the parameter posteriors evolved from ABC are much narrower, by about an order of magnitude, than with MCMC. This is understandably owed to the evolution feature in the ABC algorithm which shrinks the parameters as the competition becomes tighter with each passing generation. Nonetheless, given a model and a data set, it can be seen that the Hubble constant estimates between MCMC and ABC are consistent with each other, in the sense that the ABC result is always within reasonable MCMC confidence limits. A notable exception to this however is when the SH$0$ES $H_0$ prior is considered, which then finds the MCMC and ABC results in a mild tension.

We also see the impact of the different data sets and priors on the Hubble constant best fits. We find the MCMC estimates to be consistent with one another regardless if BAO and $H_0$ priors were considered in the analysis. This also somehow holds with the ABC in the $\Lambda$CDM and DDE models except with narrower error bars. The various $H_0$ estimates are statistically consistent with each other.

Before we end, we want to remark on two more observations. First one is regarding the $H_0$ values obtained throughout this work. Our results show that the Pantheon$+$ data together with expansion rate observations at late times caters to lower $H_0$ values consistent with the Planck estimate from the CMB. This is quite transparent in Figure \ref{fig:whiskerH0} and can be seen to be the case regardless of the statistical analysis routine, data sets, and parameter priors considered. This motivates a future cosmography analysis involving the Pantheon$+$ compilation compared with the earlier Pantheon data set. The second comment is on whether ABC accounts for overfitting. This however depends on how the ABC was applied for the model selection and parameter inferencing. To be clear, the ABC is able to compare models with any number of parameters \cite{Toni_2008}, and so the only degree of freedom that may account for overfitting in the algorithm is through the distance function. Since we have considered the chi-squared, our ABC results are not transparent to overfitting. However, a naive thought may be to consider the information criterion as a distance function. The issue with this is that while the ABC model selection now takes the number of parameters into account, it will also use the information criterion to estimate the cosmological parameters, which does not make sense. We emphasize that the takeaway from our results is that the DDE was preferred over to a constant $\Lambda$. The agreement between the MCMC and ABC estimates also show that this ABC route with the chi-squared was worth considering.

{We also want to point out that the uncertainties associated with the ABC approach turn out to be very tight in comparison to traditional MCMC analyses. This points to further work needed to further understand how to best estimate such errors associated with parameter best-fit mean values. While ABC does resolve the issue of kernel selection which is born out of the Gaussian processes method, it does still leave open the issue of uncertainties associated with mean best-fit values. For this reason, the different data sets in the results tables appear to be in tension with each other at the level of $2-3\sigma$. This is an issue we hope to address in future work on the topic.}

We welcome several future directions that our work opens up. The background cosmology is only one layer of the physical universe. It is interesting whether DDE would still be favored by ABC if perturbation data is taken into account. This also calls on theorists to single out modified gravity models that naturally provides DDE. Additionally, it is interesting to see if ABC may prefer fundamental modified gravity theories, e.g., quintessence, than standard gravity. This makes for a rather compelling test as the ABC assigns an unbiased posterior over each model, regardless of the number of parameters, and then letting the data tell which is more compatible with observations. If ABC prefers any, a modified gravity can be tested in the strong gravity regime, or other observational windows, which would be able to constrain or rule out gravitational degrees of freedom beyond the ones expected in bare general relativity. {Another important direction to consider for future work is to understand how cosmological perturbations are impacted by this model, and to use them to explore early Universe data sets such as CMB data. This is important to understand further what impact these models can have on the Hubble tension problem, among others.}

\medskip

\medskip

\section*{Acknowledgement}
\noindent JLS would like to acknowledge support from the Malta Digital Innovation Authority through the IntelliVerse grant. DG acknowledges financial support from project ANIDPFCHA/Doctorado Nacional/2019-2119188. This paper is based upon work from COST Action CA21136 {\it Addressing observational tensions in cosmology with systematics and fundamental physics} (CosmoVerse) supported by COST (European Cooperation in Science and Technology).



\end{document}